\documentclass[twocolumn,prd,superscriptaddress,preprintnumbers,nofootinbib]{revtex4-1}

\usepackage{graphicx}
\usepackage{amsmath,bm,amssymb,amsfonts,dsfont}
\usepackage[usenames,dvipsnames]{xcolor}
\usepackage[normalem]{ulem}
\usepackage{url}
\usepackage{multirow}
\usepackage[colorlinks  = true,
            linkcolor   = NavyBlue,
            urlcolor    = NavyBlue,
            citecolor   = NavyBlue,
            anchorcolor = NavyBlue]{hyperref}

\usepackage{appendix}

\usepackage[switch]{lineno}

\pdfoutput=1 % if your are submitting a pdflatex (i.e. if you have
\newcommand{\lsim}{\mathrel{\mathop{\kern 0pt \rlap
  {\raise.2ex\hbox{$<$}}}
  \lower.9ex\hbox{\kern-.190em $\sim$}}}
\newcommand{\gsim}{\mathrel{\mathop{\kern 0pt \rlap
  {\raise.2ex\hbox{$>$}}}
  \lower.9ex\hbox{\kern-.190em $\sim$}}}

\begin{document}

%\preprint{CP3-XXX}
%\preprint{CTPU-PTC-24-31, CERN-TH-2024-164}

\title{CosmiXs: Spectra of Cosmic Antideuterons from Dark Matter Interactions}
\title{Nail down the theoretical uncertainties of coalescence of $\bar{D}$ produced from dark matter}
\title{Nailing down the Theoretical Uncertainties of  $\overline{\rm D}$ Spectrum Produced from Dark Matter}
\title{Nailing down the theoretical uncertainties of $\overline{\rm D}$ spectrum produced from dark matter}

\author{Mattia Di Mauro}
\email{dimauro.mattia@gmail.com}
\affiliation{Istituto Nazionale di Fisica Nucleare, Sezione di Torino, Via P. Giuria 1, 10125 Torino, Italy}

\author{Nicolao~Fornengo}
\email{nicolao.fornengo@unito.it}
\affiliation{Department of Physics, University of Torino, Via P. Giuria 1, 10125 Torino, Italy}

\author{Adil~Jueid}
\email{adiljueid@ibs.re.kr}
\affiliation{Particle Theory and Cosmology Group, Center for Theoretical Physics of the Universe, Institute for Basic Science (IBS), Daejeon, 34126, Republic of Korea}

\author{Roberto Ruiz de Austri}
\email{rruiz@ific.uv.es}
\affiliation{Instituto de F\'{\i}sica Corpuscular, CSIC-Universitat de Val\`encia, E-46980 Paterna, Valencia, Spain}

%\author{Chiara Arina}
%\email{chiara.arina@uclouvain.be}
%\affiliation{Centre for Cosmology, Particle Physics and Phenomenology (CP3), Universit\'e Catholique de Louvain, Chemin du Cyclotron 2, 1348 Louvain-la-Neuve, Belgium}

\author{Francesca Bellini}
\email{f.bellini@unibo.it}
\affiliation{University of Bologna, Department of Physics and Astronomy A. Righi, Via Irnerio 46, Bologna, 40126, BO, Italy}

\begin{abstract}
The detection of cosmic antideuterons ($\overline{\rm D}$) at kinetic energies below a few GeV/n could provide a smoking gun signature for dark matter (DM). However, the theoretical uncertainties of coalescence models have represented so far one of the main limiting factors for precise predictions of the $\overline{\rm D}$ flux. In this Letter we present a novel calculation of the $\overline{\rm D}$ source spectra, based on the Wigner formalism, for which we implement the Argonne $v_{18}$ antideuteron wavefunction that does not have any free parameters related to the coalescence process. 
We show that the Argonne Wigner model excellently reproduces the $\overline{\rm D}$ multiplicity measured by ALEPH at the $Z$-boson pole, which is usually adopted to tune the coalescence models based on different approaches.
Our analysis is based on \textsc{Pythia}~8 Monte Carlo event generator and the state-of-the-art \textsc{Vincia} shower algorithm. We succeed, with our model, to reduce the current theoretical uncertainty on the prediction of the $\overline{\rm D}$ source spectra to a few percent, for $\overline{\rm D}$ kinetic energies relevant to DM searches with GAPS and AMS, and for DM masses above a few tens of GeV. This result implies that the theoretical uncertainties due to the coalescence process are no longer the main limiting factor in the predictions. We provide the tabulated  source spectra for all the relevant DM annihilation/decay channels and DM masses between 5 GeV and 100 TeV, on the \href{https://github.com/ajueid/CosmiXs.git}{CosmiXs github repository}.
\end{abstract}

%\keywords{Dark Matter, Indirect Detection experiments, Monte Carlo event generators.}
%{\let\thefootnote\relax
%\footnotetext{$^\dagger$\,Contact authors} }

\maketitle

\flushbottom

{\it Introduction} --
Despite decades of theoretical studies and experimental searches, the particle origin of dark matter (DM) has eluded conclusive results. Well-motivated DM particle physics models have driven a wide search program through indirect, direct detection and collider experiments. In indirect detection, researchers seek possible signatures of DM signals in the flux of cosmic messenger particles, such as positrons, antiprotons, $\gamma$ rays, and neutrinos \cite{Gaskins:2016cha,Cirelli:2024ssz}. However, identifying DM through the detection of these particles is challenging as the majority of their fluxes is likely to be attributed to known astrophysical processes  
(see, e.g., \cite{DiMauro:2015jxa,DiMauro:2015tfa,Genolini:2021doh,DiMauro:2021qcf,McDaniel:2023bju,Balan:2023lwg}). 

Cosmic antinuclei originating from DM annihilation or decay might instead offer a more advantageous option. In particular, cosmic antideuterons ($\overline{\rm D}$) \cite{Donato:1999gy}  -- and, to some extent, antihelions ($\overline{\rm He}$) \cite{Cirelli:2014qia, Carlson:2014ssa} -- have emerged as particularly promising channels for DM indirect detection, largely due to the suppressed low-energy flux from astrophysical sources and secondary production. 
In the context of Weakly Interacting Massive Particles (WIMPs) annihilating within the Galactic halo, the flux of $\overline{\rm D}$ can be at least one order of magnitude higher than that of secondary antinuclei within kinetic energy ranges of 0.1--1 GeV/nucleon (see, e.g., \cite{Donato:1999gy,Ibarra:2012cc,Fornengo:2013osa,Herms:2016vop,Korsmeier:2017xzj}). As a result, even the detection of a small number of $\overline{\rm D}$ events at these energies could serve as a compelling signature of DM \cite{vonDoetinchem:2015yva}. In particular, the AMS-02 experiment \cite{2008ICRC....4..765C} and the future GAPS observational campaigns could have the sensitivity to detect a few $\overline{\rm D}$ events \cite{Aramaki:2015laa}.

Even though DM indirect detection through antinuclei leads potentially to the clearest signature among all the possible cosmic messengers, the theoretical framework for their production is not fully understood. Many studies on this topic employ the so-called {\it coalescence models} \cite{Kapusta:1980, Butler:1963, Scheibl:1998tk}, which posit that individual antinucleons (antiprotons and antineutrons) bind into antinuclei if the difference in momentum between them is smaller than a coalescence momentum  -- denoted hereafter by $p_{\rm{coal}}$ --. Unfortunately, the value of $p_{\rm{coal}}$ itself cannot be determined from first principles and may vary depending on the process generating the nucleus, {\it i.e.} whether from DM annihilation or secondary (hadronic) production. 
For instance, the authors of Refs.~\cite{Ibarra:2012cc,Fornengo:2013osa,Korsmeier:2017xzj,Kachelriess:2020uoh} 
have found $p_{\rm{coal}}\approx 0.15$ GeV when calibrating the DM annihilation processes using the rates of $\overline{\rm D}$ production at the $Z$ resonance measured by the Large Electron Positron (LEP) collider reported on by ALEPH and OPAL collaborations \cite{2006192,1995ZPhyC..67..203A}.
Conversely, when using ALICE data for $p-p$ collisions with center-of-mass energies ranging from 900 GeV to 13 TeV \cite{ALICE:2015wav,ALICE:2017xrp,ALICE:2019dgz,ALICE:2020foi,ALICE:2021mfm,ALICE:2021ovi} to calibrate $\overline{\rm D}$ secondary production, values $p_{\rm{coal}}>0.2$ GeV are found.
As the $\overline{\rm D}$ spectra are roughly proportional to the cube of $p_{\rm{coal}}$, a variation of this parameter between 0.15 and 0.22, as found in the literature, would induce a theoretical uncertainty of approximately a factor of 3.

More sophisticated coalescence models, based on a quantum-mechanical treatment of the problem using a Wigner formalism, consider that the process depends on the momentum distribution of the nucleons, the nucleus wave function, and the characteristics of the nucleon emitting source~\cite{Scheibl:1998tk,Bellini:2018epz, Blum:2017qnn, Bellini:2020cbj, Kachelriess:2019taq,Mahlein:2023fmx}. This approach is the foundation of most recent developments that calculate the production of (anti)nuclei in hadronic interactions on an event-by-event basis by employing Monte Carlo (MC) simulations~\cite{Kachelriess:2019taq,Kachelriess:2020uoh,Mahlein:2023fmx}. By using the Wigner approach, these models can account for both the momentum and spatial correlations of antiprotons ($\bar{p}$) and antineutrons ($\bar{n}$) that contribute to the production of antinuclei.
However, even if this treatment is more physically motivated, in literature the choice of the nucleons and the $\overline{\rm D}$ wave function falls on simplified models such as single or double Gaussian functions, whose parameters are tuned on nuclei production data, (see, e.g., Refs.~\cite{Kachelriess:2019taq,Kachelriess:2020uoh}).

In this Letter, we aim to contribute to the improvement of the predictions of the $\overline{\rm D}$ {\sl source spectra} produced from DM annihilation or decay. To this aim, we propose the adoption of a model which implements the Argonne nucleon-nucleon potential, with explicit charge dependence and charge asymmetry \cite{Wiringa:1994wb}, in a Wigner formalism (hereafter {\sl Argonne Wigner model}) for the full quantum-mechanical treatment of the coalescence process. 
The Argonne potential is tuned directly to $p$--$p$ and $n$--$p$ inelastic scattering data, low-energy $n$--$n$ scattering processes, and to the deuteron binding energy \cite{Wiringa:1994wb}. This approach, which was never used before for the calculations of $\overline{\rm D}$ spectra from DM yields, it is more predictive than other approaches since the Argonne nucleon-nucleon potential is completely fixed by $n$, $p$ scattering data and thus the model does not have free unknown parameter related to the coalescence mechanism, while at the same time yielding excellent agreement with the ALEPH measurement  \cite{ALEPH:2006qoi} of the $\overline{\rm D}$ multiplicity \footnote{Note that the validity of the adoption of an Argonne Wigner approach for the production of (anti)deuterons was also demonstrated in Ref.~\cite{Mahlein:2023fmx} for {\sl proton-proton} collisions at $\sqrt{s} = 13$ TeV compared to ALICE data, in a semi-analytical approach.}.  We implement the Argonne Wigner model in a fully MC-based approach where we generate DM particle annihilations, search for every pair of $\bar{n}$ and $\bar{p}$ produced in the annihilation process and decide whether a $\overline{\rm D}$ is formed on the basis of the Argonne Wigner formalism, described below. Employing an MC generator enables us to properly take into account both spatial and momentum correlations between antinucleons. For this, we use \textsc{Pythia}~8 \cite{Bierlich:2022pfr} with the \textsc{Vincia} shower algorithm \cite{Fischer:2016vfv}, which accurately incorporates various relevant effects, such as electroweak corrections and effects related to the spin of the particles involved in the processes (see Ref.~\cite{Arina:2023eic} and Appendix \ref{app:wigner} for all the details and the relevance of using \textsc{Vincia}). We publicly distribute the source spectra of $\overline{\rm D}$, for quark and boson annihilation channels and for DM masses between 5 GeV and 100 TeV and for all the models considered in this Letter, on the CosmiXs github repository\footnote{\href{https://github.com/ajueid/CosmiXs.git}{https://github.com/ajueid/CosmiXs.git}}, as part of the CosmiXs suite \cite{Arina:2023eic}. We do not consider leptonic channels because their contribution to the $\overline{\rm D}$ yield is negligible.

\begin{figure*}
\includegraphics[width=0.49\linewidth]{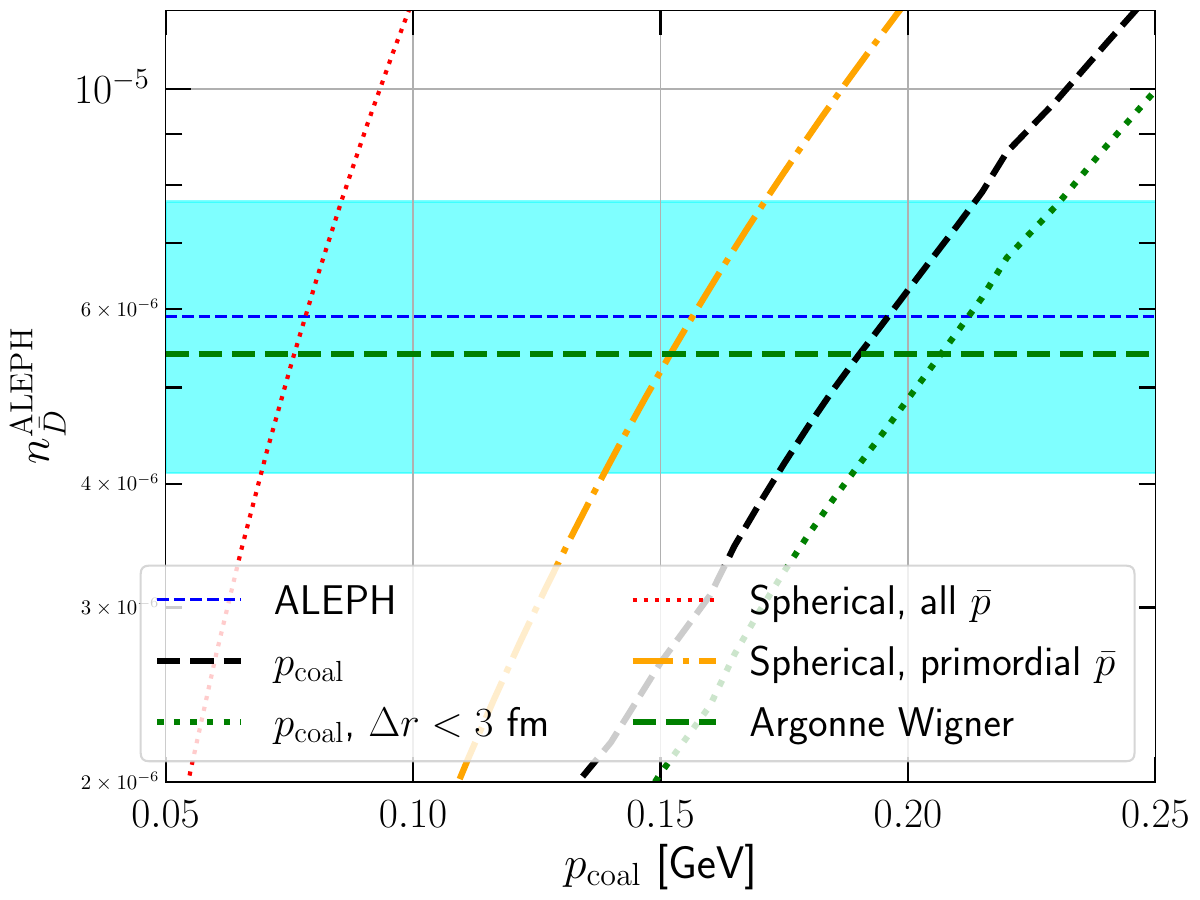}
\includegraphics[width=0.49\linewidth]{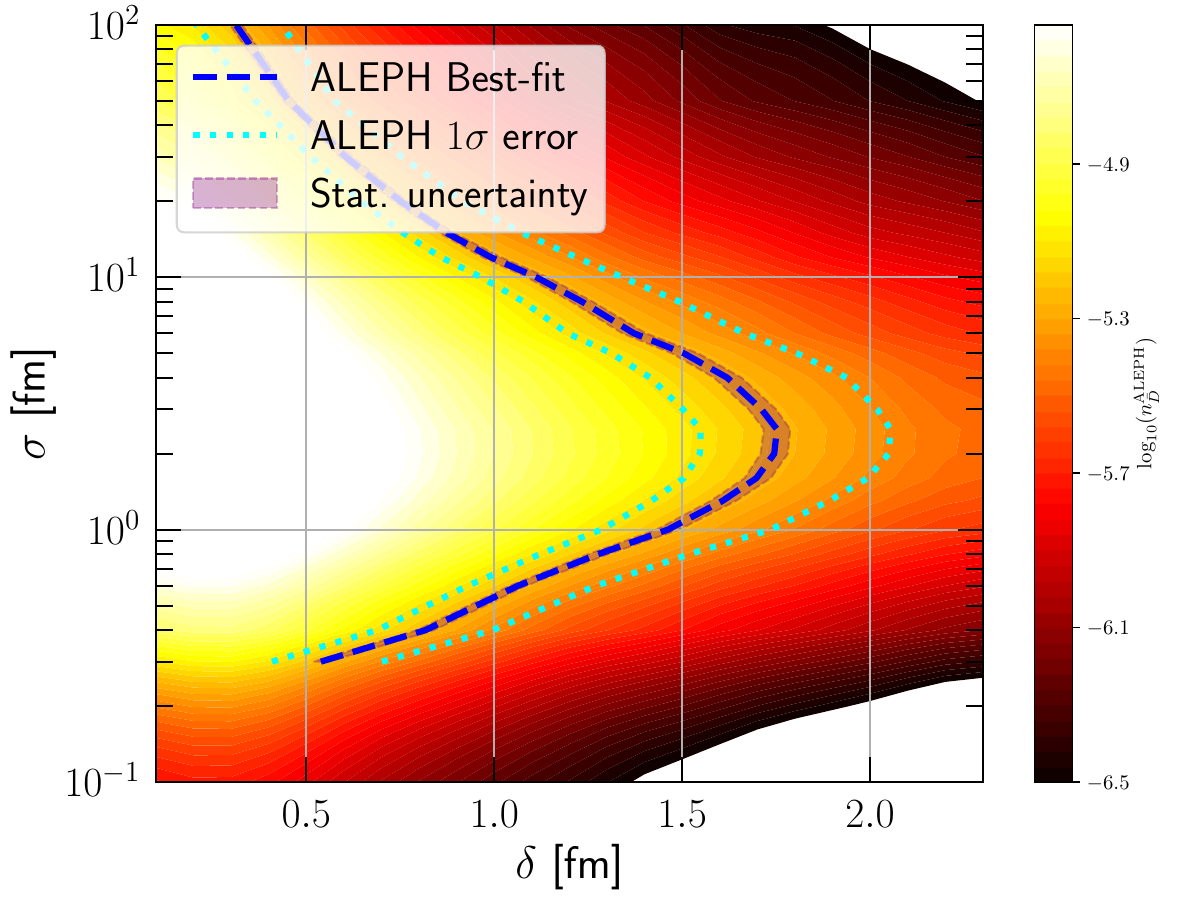}
    \caption{Left Panel: Variation of the theoretical prediction for the $\overline{\rm D}$ multiplicity as a function of the parameter $p_{\rm{coal}}$ compared with the ALEPH data.  
    The other curves represent the result obtained with the various coalescence models.
    For the spherical model we show the result obtained when considering all the antinucleons or only the primordial ones. Right Panel: Contour plot of the $\overline{\rm D}$ multiplicity obtained for different combination of the parameters $\sigma$ and $\delta$ for the Gauss Wigner model. The dashed blue curve and greed dotted curves represent the best-fit and its $1\sigma$ uncertainty for the ALEPH data, while the pink band represents the statistical errors due to the number of events we have simulated with \textsc{Pythia}.}
    \label{fig:coalparams}
\end{figure*}

{\it Coalescence models} --
The coalescence models that we adopt and compare in our analyses are outlined below:
\begin{enumerate}
\item {\it Simple coalescence model} (labeled as $p_{\rm{coal}}$): We select, for each annihilation event in the MC simulations, the $\bar{n}$ and $\bar{p}$ pairs that have a difference in momentum $\Delta p = |\vec{p}_{\bar{p}} - \vec{p}_{\bar{n}}|$, in the $\overline{\rm D}$ rest frame, smaller than $p_{\rm{coal}}$, which is the single parameter of the model. We select only {\it primordial} antinucleons coming from hadronization or from resonance decays of heavy baryons. On the contrary, we exclude antinucleons generated from weak decays of different hadrons because they would be significantly displaced in space-time and therefore unable to coalesce.
\item {\it Simple coalescence model with a sharp cutoff in distance} ($p_{\rm{coal}}$, $\Delta r<3$ fm). This model generalizes the previous one with the addition of a sharp cut on the distance $\Delta r$ between $\bar{n}$ and $\bar{p}$, which should be, in the rest frame of the $\overline{\rm D}$, smaller than 3 fm \footnote{This assumption is motivated by the size of the wavefunction of the harmonic oscillator potential $d$, which, if taken to be $d=3.2$ fm, reproduces the measured deuteron root-mean-square (RMS) charge radius \cite{CREMA:2016idx}.}. 
\item {\it Wigner approach with a Gaussian wavefunction (Gauss Wigner).} This model implements the Wigner formalism, with the assumption that the deuteron internal wavefunction is modeled by a Gaussian distribution in space and momentum:
\begin{equation}
     \mathcal{D}(\Delta r, \Delta p) \propto e^{-\Delta r^2/(2 \sigma^2)}e^{-\Delta p^2 \delta^2/2},
     \label{eq:GWF}
\end{equation}
where $\sigma$ and $1/\delta$ account for the Gaussian distribution widths for the antinucleons separations in space and momentum. Following the Wigner formalism (see Appendix \ref{app:wigner}), these widths should be of about the same order as the deuteron size, which is about $d/\sqrt{2} \approx 2$ fm. We implement the Wigner model making the assumption that the function $\mathcal{D}(\Delta r, \Delta p)$ is a Probability Distribution Function (PDF) for $\Delta r$ and $\Delta p$. Therefore, for every pair of $\bar{n}$ and $\bar{p}$ we evaluate $\mathcal{D}(\Delta r, \Delta p)$, we draw a random number between $[0,1]$ and if this number is smaller than the value of $\mathcal{D}(\Delta r, \Delta p)$, the $\overline{\rm D}$ is formed.
\item {\it Wigner approach with Argonne function (Argonne Wigner).}  This method follows the same approach as the previous one but it uses the Argonne $v_{18}$ function \cite{Wiringa:1994wb}. This model is entirely fixed and has no free parameters. In particular, we have tabulated $\mathcal{D}(\Delta r, \Delta p)$ following the prescriptions of Ref.~\cite{Mahlein:2023fmx}, which provides an analytic expression for the $S$ and $D$-wave components of the Wigner function. This is our reference model.
\item {\it Spherical approach}. This model is the simplest (and oldest) one and assumes that all the pairs of $\bar{p}$ and $\bar{n}$ are uncorrelated in space and momentum and are produced with a spherical symmetry in their center-of-mass frame. With these approximations, the spectra of $\overline{\rm D}$ is proportional to the square of the nucleons spectra and to the coalescence momentum to the third power \cite{Donato:1999gy,Fornengo:2013osa,Korsmeier:2017xzj}. 
This model is based on effective but simple assumptions that are actually not suitable for DM annihilation channels into gauge and Higgs bosons, and it typically underestimates the high-energy part of the spectra arising from quark hadronization \cite{Fornengo:2013osa}. Nevertheless, we include it in the list for the sake of comparison.
\end{enumerate}
We explain in more detail the models using the Wigner formalism as well the {\it Simple coalescence models} and their application into the MC framework in the Appendices \ref{sec:argonne} and \ref{sec:montecarlo}.

{\it Coalescence models tuning} --
We first tune all models (except Argonne Wigner, which has no free parameter) to the $\overline{\rm D}$ multiplicity measured by ALEPH \cite{2006192} for the $e^+e^-$ process at the $Z$-pole: $n^{\rm{ALEPH}}_{\overline{\rm D}} = (5.9 \pm 1.8 \pm 0.5) \times 10^{-6}$ per hadronic $Z$ decay in the $\overline{\rm D}$ momentum range (0.62, 1.03) GeV/c and for polar angles $|\cos{\theta}| < 0.95$.
In Fig.~\ref{fig:coalparams} we show the variation of the $\overline{\rm D}$ multiplicity for the different models. 
For the spherical and $p_{\rm coal}$ models, it is crucial to consider only primordial antinuclei, which are the only ones able to form an antideuteron detectable by ALEPH. By including also antinucleons produced from weak decays, we would get a best-fit of $p_0$ which is a factor of 3 smaller, because the antinucleons yield would be incorrectly enhanced by about $40\%$.
Moreover, we see that the spherical approach gives a best fit of $p_{\rm{coal}}=0.157^{+0.015}_{-0.018}$ GeV which is typically smaller than the one obtained for simple coalescence model ($p_{\rm{coal}}=0.196^{+0.018}_{-0.023}$ GeV) where the fusion process is calculated on an event-by-event basis. We also notice that the addition of a sharp cutoff in distance between the antinucleons has some relevance, since the required value of $p_{\rm{coal}}$ to match the measured multiplicity is larger $p_{\rm{coal}}=0.196^{+0.018}_{-0.023}$ GeV. This is due to the fact that without adding the condition that the $\bar{p}$ and $\bar{n}$ must be close in space for a $\overline{\rm D}$ to form, a larger number of antinucleons are allowed to coalesce, some of which unphysically. 

\begin{figure*}
\includegraphics[width=0.49\linewidth]{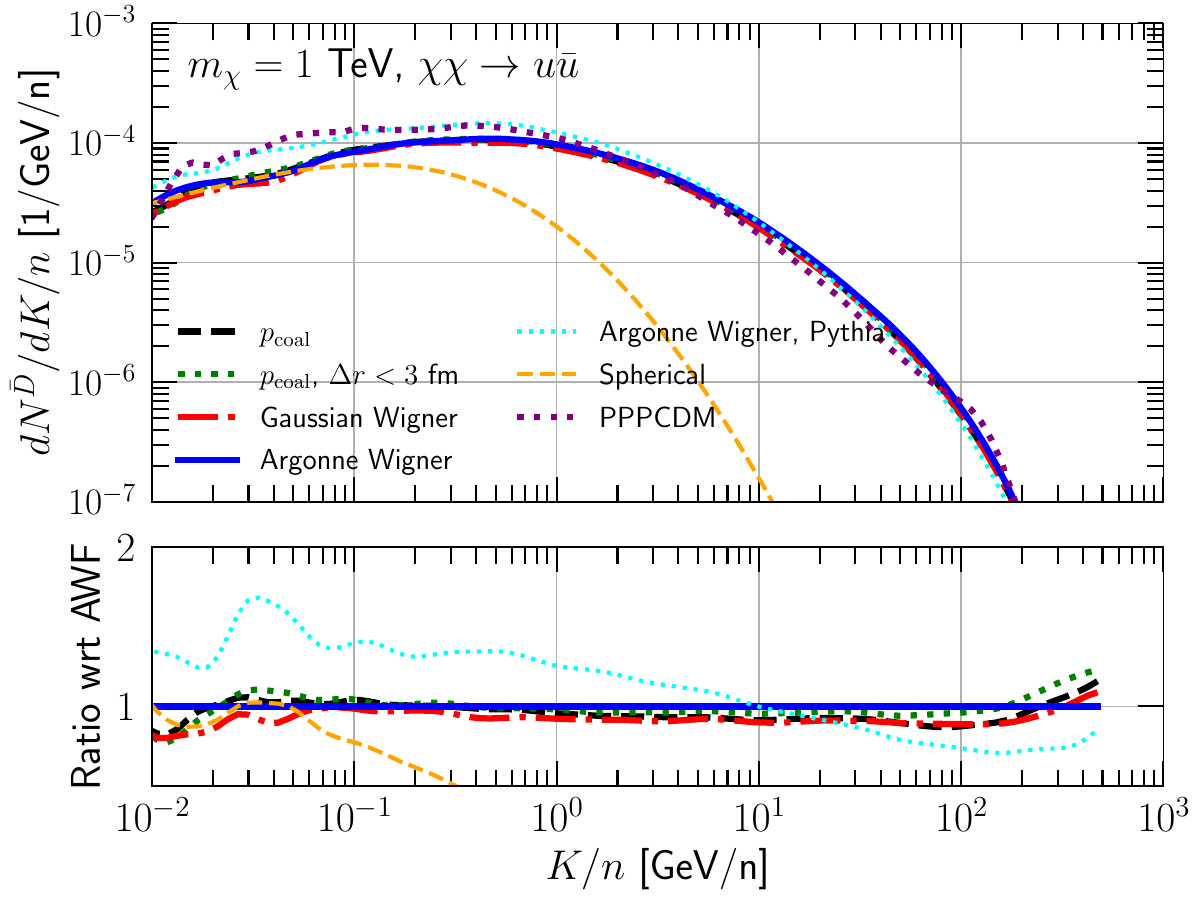}
\includegraphics[width=0.49\linewidth]{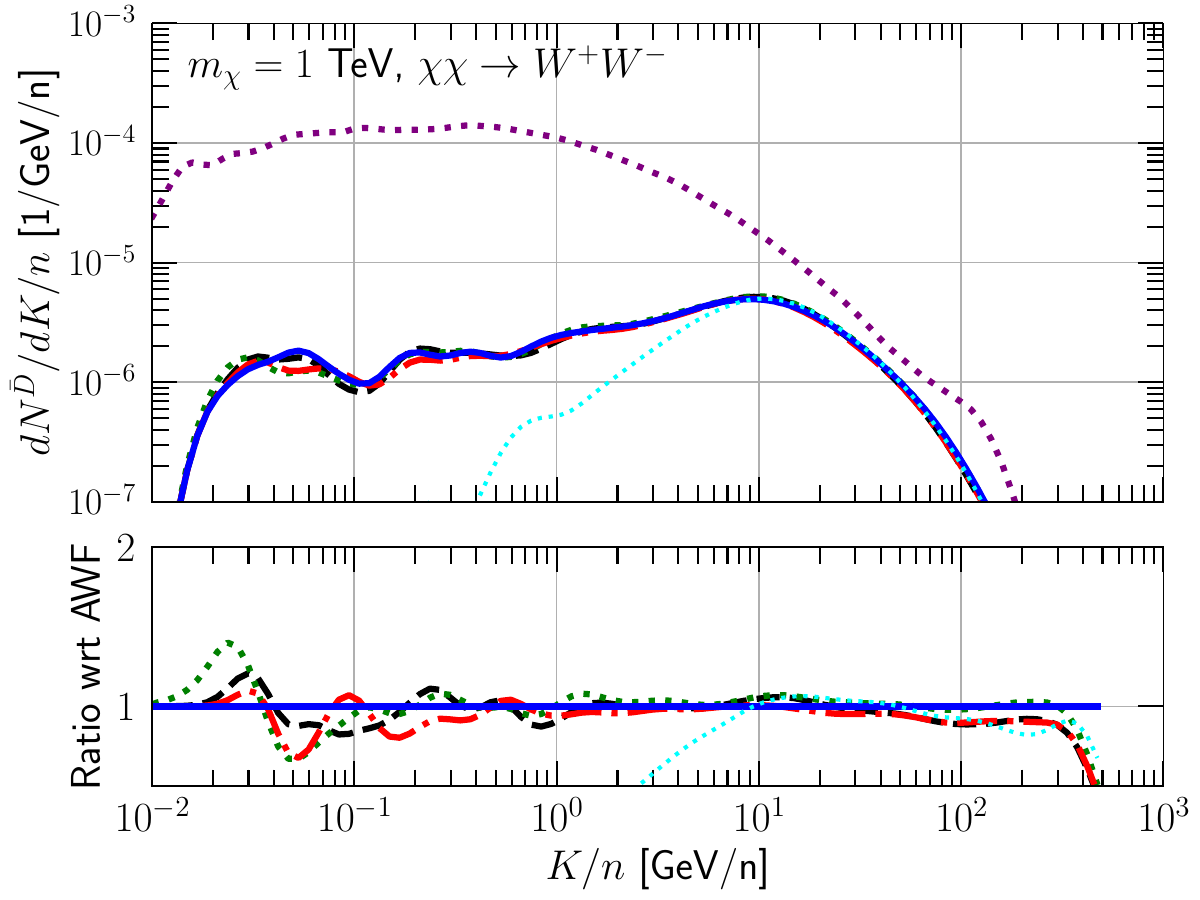}
    \caption{Spectra of $\overline{\rm D}$ as a function of kinetic energy per nucleon, for the $u\bar{u}$ (left panel) and $W^+W^-$ (right panel) annihilation channels and $m_{\rm{DM}}= 1$ TeV. The case denoted Argonne Wigner Pythia refers to the adoption of the standard \textsc{Pythia} shower algorithm instead of \textsc{Vincia}. For comparison, we also display the spectra calculated with PPPCDM. The lower panels show the ratio between the spectra obtained in different models and the ones obtained with the Argonne Wigner. \cite{Cirelli:2010xx}.}
    \label{fig:spectrum}
\end{figure*}

The Gauss Wigner model instead has two parameters, $\sigma$ and $\delta$, which exhibit some level of degeneracy, as can be seen in the right panel of Fig.~\ref{fig:coalparams}. The two parameters are positively correlated starting from small values of both $\delta$ and $\sigma$, and then negatively correlated after a turnaround of the best-fit curve occurring for the spatial dispersion parameter of the Gaussian $\sigma\approx 2$--$3$ fm, a value compatible with the $\overline{\rm D}$ size $d\approx 3$ fm. In the Wigner formalism, discussed in Appendix \ref{app:wigner}, the parameters $\delta$ and $\sigma$ are actually restricted to take the same value, which for the $\overline{\rm D}$ is of the order of $d/\sqrt{2}\approx 2$ fm. Therefore, we adopt hereafter the condition of equality of the two parameters and identify as $\delta = \sigma = 1.8$ fm the point along the best-fit curve in Fig. \ref{fig:coalparams}. We obtain a $\overline{\rm D}$ multiplicity of $5.4 \times 10^{-6}$ with the Argonne Wigner (see Fig.~\ref{fig:coalparams}), in excellent agreement with the ALEPH data within its $1\sigma$ error.

{\it $\overline{\rm D}$ spectra}--
A sample of the $\overline{\rm D}$ spectra from DM annihilation, calculated for the various
coalescence models and obtained using \textsc{Pythia}~8.309 with the \textsc{Vincia} shower algorithm, is presented in Fig.~\ref{fig:spectrum}.
The first relevant result is that, once properly tuned, the spectra obtained with different coalescence models are very similar, differing at most about $10$--$20\%$ among each other. The largest variations occur at the high-energy endpoint of the spectra, close to the kinematical cutoff resulting from the energy approaching the value of the DM mass and, to some extent, at the lowest energies, which are nevertheless below the reach of current and foreseen experiments like AMS-02 and GAPS. For a wide range of energies, the spectra obtained with the various models agree at the few percent level, and this occurs for all the DM annihilation final states. 
Fig.~\ref{fig:spectrum}
also shows that the Argonne Wigner model produces remarkably consistent results, thus making the theoretical predictions of the $\overline{\rm D}$ spectra robust.

We display in Fig.~\ref{fig:spectrum} the differences between the adoption of the \textsc{Vincia} showering algorithm in \textsc{Pythia} as compared to the standard and \textsc{Pythia}, which, for the quark channels, can reach $40$--$50\%$ around the peak of the spectrum. In the case of the gauge boson channels, this difference is even higher because the effects of the electroweak corrections are extremely important. We also notice that the results we obtain for the quark channels are similar to the ones obtained in the PPPCDM \cite{Cirelli:2010xx} while instead, for the gauge bosons channels our predictions are quite different from the one from PPPCDM at low energies. This could be due to the differences in the treatment of electroweak corrections which have already been shown and discussed in Ref. \cite{Arina:2023eic}.

A clear way of appreciating the variation between the different coalescence models is given by the integrated multiplicity of $\overline{\rm D}$ per annihilation event. This is shown as a function of the DM mass in Fig.~\ref{fig:mult} for the $b \bar{b}$ and $W^+W^-$ channels and for DM masses between 1 and $10^5$ GeV. Apart from the spherical model, which was already shown to be incorrect and is here reported only for completeness, we see that the various models predict a very similar multiplicity, with differences which are confined to less than 10\% for all DM masses and channels, except for the $b \bar{b}$ channel when the mass of the DM is close to the center-of-mass energy (i.e.~for DM masses close to 10 GeV) that allows the formation of $b$-baryons and mesons. 
As already noted for the source spectra (see Fig.~\ref{fig:spectrum}), we find also for the $\overline{\rm D}$ multiplicities in Fig.~\ref{fig:mult} discrepancies between our results for the $W^+W^-$ and the one in Ref.~\cite{Cirelli:2010xx}. While the $\overline{\rm D}$ is flat for increasing DM masses, as previously shown also in Refs.~\cite{Kadastik:2009ts,Fornengo:2013osa}, Ref.~\cite{Cirelli:2010xx} predicts an increasing shape.

\begin{figure}
\includegraphics[width=0.99\linewidth]{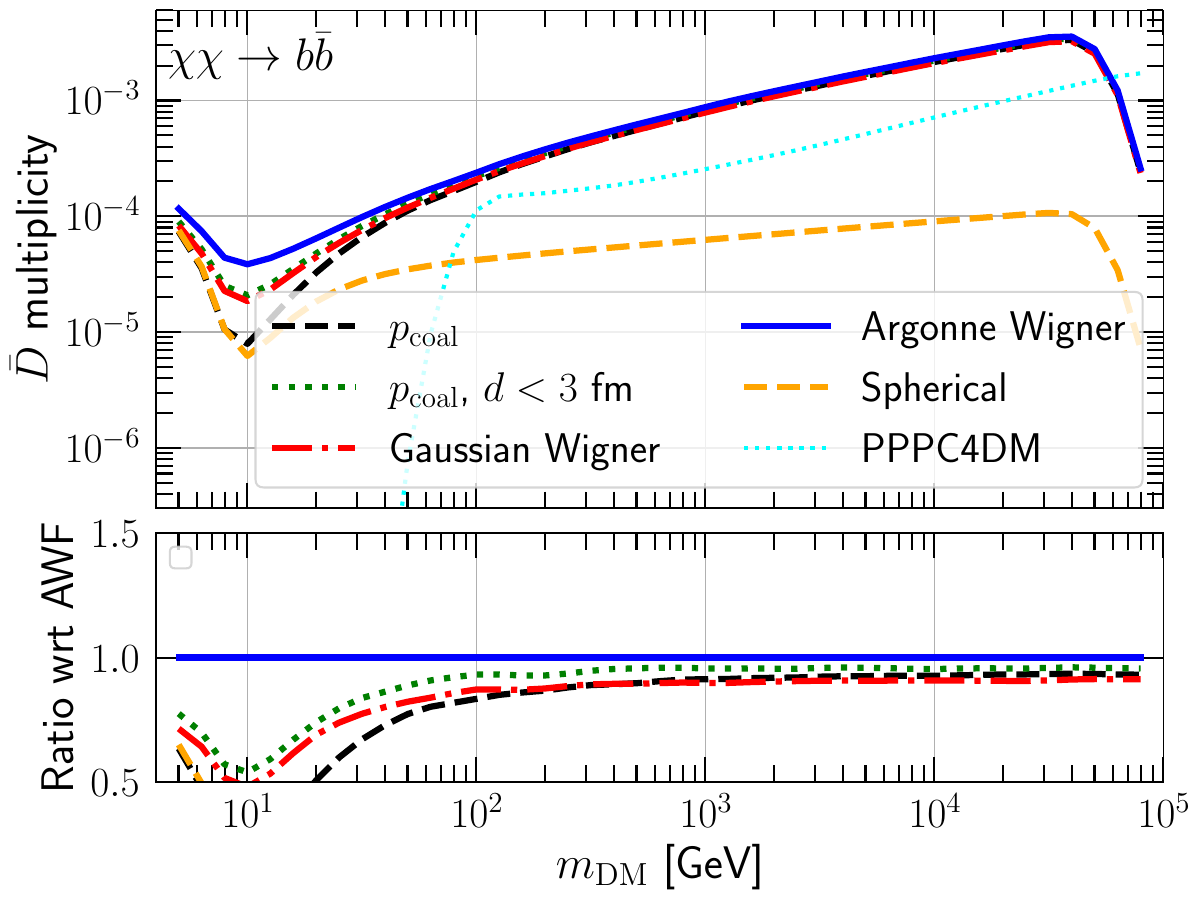}
\includegraphics[width=0.99\linewidth]{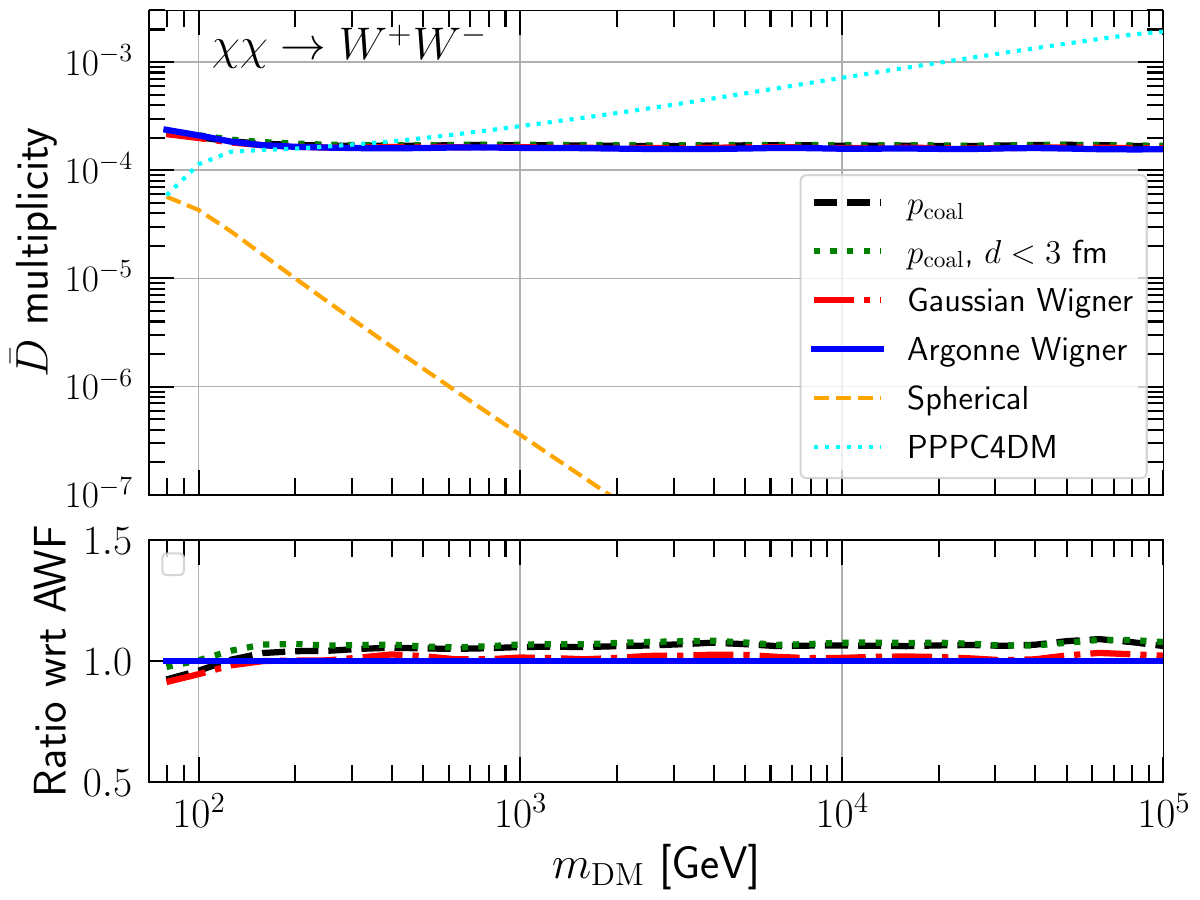}
    \caption{$\overline{\rm D}$ multiplicity per annihilation event, calculated for the $b \bar{b}$ and $W^+W^-$ annihilation channels and DM masses between 1 and $10^5$ GeV and obtained for the different coalescence models discussed in the paper. Notation as in Fig. \ref{fig:spectrum}.}
    \label{fig:mult}
\end{figure}

\begin{figure}
\includegraphics[width=0.99\linewidth]{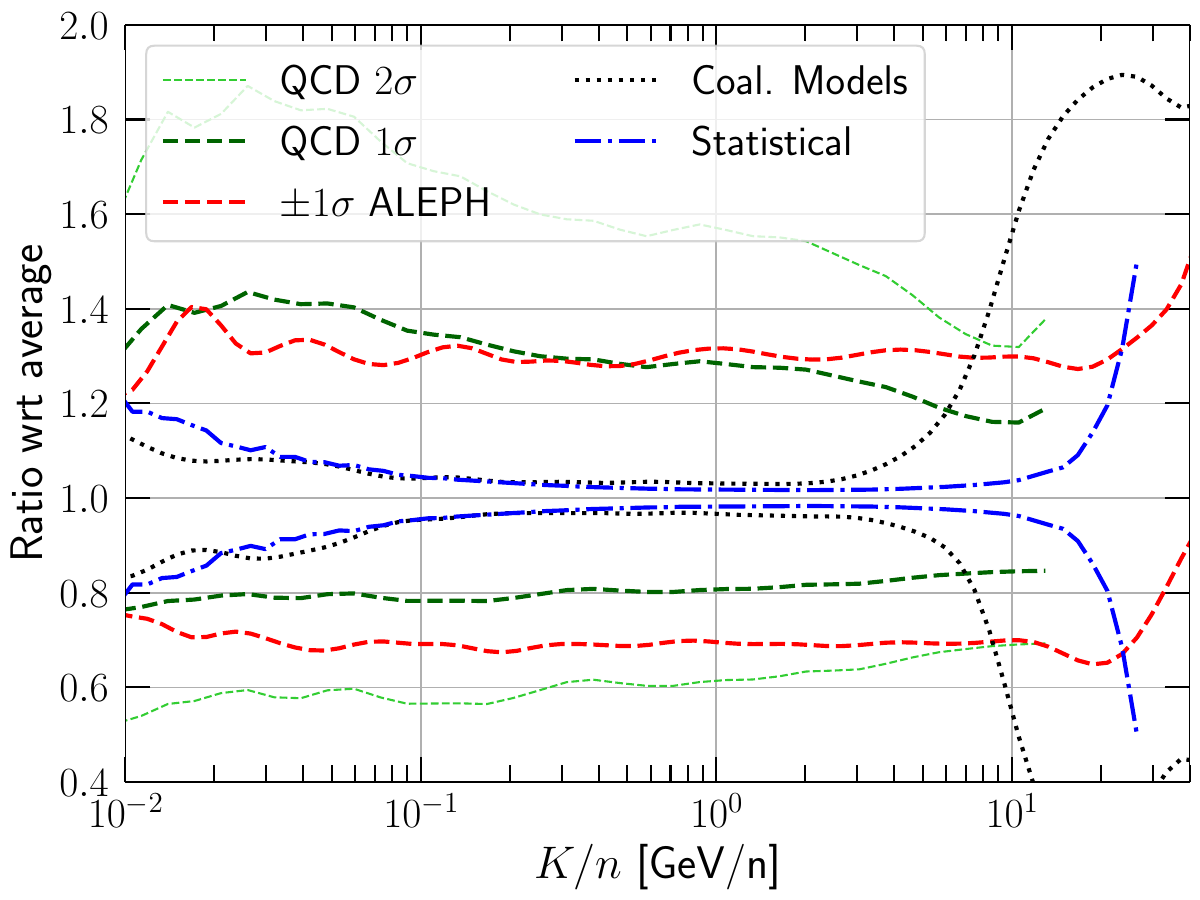}
    \caption{Uncertainties on the $\overline{\rm D}$ spectra arising from the fit to the ALEPH data (region inside the red curves), statistical uncertainties from the MC sampling (blue curves), maximal differences between the coalescence models (black) and uncertainties from hadronization (green). The results are shown as a function of the kinetic energy per nucleon and for the case of the $b\bar{b}$ annihilation channel for $m_{\rm DM} = 100$ GeV.}
    \label{fig:unc}
\end{figure}

{\it Uncertainties} --
We now turn into a discussion of the different sources of uncertainties that may affect the $\overline{\rm D}$ spectra. To be definite, we show in Fig.~\ref{fig:unc} the results obtained for the $b\bar{b}$ channel, $m_{\rm{DM}}=100$ GeV and the Argonne Wigner model, but the same results apply in general.

First of all, there is a statistical uncertainty related to the limitation in the size of the MC samples we have generated and the ensuing $\overline{\rm D}$ multiplicity for each kinetic energy bin. In this study, we have generated a number of events that depends on the DM mass in order to reach a statistical uncertainty of the order of $5\%$ at the peak of the spectra and less than 10\% for the whole kinetic-energy range of interest for the current and foreseen detectors (AMS-02 and GAPS). For example, with a DM mass of 100 GeV we have simulated 200 million events.

The second source of uncertainties arises from the choice of the coalescence model. We observe that, when properly calibrated on $\overline{\rm D}$ production data, all models produce very consistent results: the span of the predictions of the various model is confined to be of the order of 5\%--10\% in the most relevant kinematic region for DM searches. This is true also when comparing the source spectra of these models with the predictions obtained with the parameter-free Argonne Wigner model. The consistency of the $\overline{\rm D}$ spectra predictions among different approaches has not been appreciated in the literature, where variations up to a factor of a few have been reported: we can therefore conclude that the choice of the coalescence model is not a limiting factor for accurate determinations of the $\overline{\rm D}$ source spectra.

For models based on a tuning on the ALEPH data, the uncertainty on the measured $\overline{\rm D}$ multiplicity translates into a 30\% variation on the predicted spectra. The uncertainty of the Argonne Wigner wavefunction induced by the fit to the experimental errors on the proton-neutron scattering measurements has never been robustly estimated \cite{Wiringa:1994wb}. However, these scattering data have errors typically smaller than $10\%$ \cite{PhysRevC.48.792}. Therefore, we can conclude that the uncertainty of the Argonne wavefunction induced by the fit should be smaller than $10\%$ and therefore the adoption of this model allows for a reduction of the theoretical uncertainty in the predicted $\overline{\rm D}$ fluxes.

Finally, the $\bar p$ and $\bar n$ production through MC event generators, which is common to all $\overline{\rm D}$ formation models, have QCD uncertainties related to the parametric variation of the Lund hadronization model parameters (see Refs. \cite{Jueid:2022qjg,Jueid:2023vrb}) as the dominant source of errors. These uncertainties are quantified at the level of $20\%$ for the $1\sigma$ eigentunes. For the more conservative scenario where $2\sigma$ eigentunes are adopted, uncertainies on the $\overline{\rm D}$ spectra can reach the level of 50\%--60\%.

{\it Conclusions}--
In this letter, we have considered the Argonne $v_{18}$ function in a Wigner formalism for the first time to calculate the source spectra of $\overline{\rm D}$ from DM annihilation. This is the most physically motivated model among the ones considered, because it is calibrated on a rich $p–p$ and $n–p$ inelastic scattering dataset, thus making it highly predictive for DM applications also for the energies and DM masses different from the ALEPH $\overline{\rm D}$  multiplicity data.
We also showed that the Argonne Wigner model properly reproduces this observable, without further specific tuning. The tabulated spectra derived with the Argonne function, that we distribute on the \href{https://github.com/ajueid/CosmiXs.git}{CosmiXs github repository}, should thus be considered as the state-of-the-art for the $\overline{\rm D}$ spectra from DM annihilation. We also showed that the spectra of $\overline{\rm D}$ produced from DM annihilation obtained with simple coalescence models and calculations based on the Wigner formalism are compatible within a 10\% uncertainty, provided that the coalescence models (other than the Argonne one) are properly calibrated to the ALEPH measurement of the $\overline{\rm D}$ multiplicity. 
We can therefore conclude that the theoretical uncertainties arising from the process of $\overline{\rm D}$ formation are no longer a limitation for the prediction for DM studies, since they all converge to a common prediction, with a small spread.

\begin{acknowledgments}
We thank Christian Bierlich, Peter Skands and all the Pythia team for the help provided with the code.
We also thank Marco Cirelli for insightful discussions and comments.
N.F.~and M.D.M.~acknowledge support from the research grant {\sl TAsP (Theoretical Astroparticle Physics)} funded by Istituto Nazionale di Fisica Nucleare (INFN). The work of A.J. is supported by the Institute for Basic Science (IBS) under the project code, IBS-R018-D1. A.J. would like to thank the CERN Theoretical Physics department for its hospitality where part of this work has been done. R.R.dA. acknowledges support from the Ministerio de Ciencia y Innovación (PID2020-113644GB-I00) and the GVA Research Project {\sl Sabor y Origen de la Materia (SOM)} (PROMETEO/2022/069). F.B. acknowledges support from the European Research Council under the European Union’s Horizon 2020 research and innovation programme through the ERC-H2020-STG CosmicAntiNuclei project (GA n. 950692).
\end{acknowledgments}

%\newpage

%\appendix

\bibliographystyle{apsrev4-1}
\bibliography{main.bib}

\begin{thebibliography}{53}%
\makeatletter
\providecommand \@ifxundefined [1]{%
 \@ifx{#1\undefined}
}%
\providecommand \@ifnum [1]{%
 \ifnum #1\expandafter \@firstoftwo
 \else \expandafter \@secondoftwo
 \fi
}%
\providecommand \@ifx [1]{%
 \ifx #1\expandafter \@firstoftwo
 \else \expandafter \@secondoftwo
 \fi
}%
\providecommand \natexlab [1]{#1}%
\providecommand \enquote  [1]{``#1''}%
\providecommand \bibnamefont  [1]{#1}%
\providecommand \bibfnamefont [1]{#1}%
\providecommand \citenamefont [1]{#1}%
\providecommand \href@noop [0]{\@secondoftwo}%
\providecommand \href [0]{\begingroup \@sanitize@url \@href}%
\providecommand \@href[1]{\@@startlink{#1}\@@href}%
\providecommand \@@href[1]{\endgroup#1\@@endlink}%
\providecommand \@sanitize@url [0]{\catcode `\\12\catcode `\$12\catcode
  `\&12\catcode `\#12\catcode `\^12\catcode `\_12\catcode `\%12\relax}%
\providecommand \@@startlink[1]{}%
\providecommand \@@endlink[0]{}%
\providecommand \url  [0]{\begingroup\@sanitize@url \@url }%
\providecommand \@url [1]{\endgroup\@href {#1}{\urlprefix }}%
\providecommand \urlprefix  [0]{URL }%
\providecommand \Eprint [0]{\href }%
\providecommand \doibase [0]{http://dx.doi.org/}%
\providecommand \selectlanguage [0]{\@gobble}%
\providecommand \bibinfo  [0]{\@secondoftwo}%
\providecommand \bibfield  [0]{\@secondoftwo}%
\providecommand \translation [1]{[#1]}%
\providecommand \BibitemOpen [0]{}%
\providecommand \bibitemStop [0]{}%
\providecommand \bibitemNoStop [0]{.\EOS\space}%
\providecommand \EOS [0]{\spacefactor3000\relax}%
\providecommand \BibitemShut  [1]{\csname bibitem#1\endcsname}%
\let\auto@bib@innerbib\@empty
%</preamble>
\bibitem [{\citenamefont {Gaskins}(2016)}]{Gaskins:2016cha}%
  \BibitemOpen
  \bibfield  {author} {\bibinfo {author} {\bibfnamefont {J.~M.}\ \bibnamefont
  {Gaskins}},\ }\href {\doibase 10.1080/00107514.2016.1175160} {\bibfield
  {journal} {\bibinfo  {journal} {Contemp. Phys.}\ }\textbf {\bibinfo {volume}
  {57}},\ \bibinfo {pages} {496} (\bibinfo {year} {2016})},\ \Eprint
  {http://arxiv.org/abs/1604.00014} {arXiv:1604.00014 [astro-ph.HE]}
  \BibitemShut {NoStop}%
\bibitem [{\citenamefont {Cirelli}\ \emph {et~al.}(2024)\citenamefont
  {Cirelli}, \citenamefont {Strumia},\ and\ \citenamefont
  {Zupan}}]{Cirelli:2024ssz}%
  \BibitemOpen
  \bibfield  {author} {\bibinfo {author} {\bibfnamefont {M.}~\bibnamefont
  {Cirelli}}, \bibinfo {author} {\bibfnamefont {A.}~\bibnamefont {Strumia}}, \
  and\ \bibinfo {author} {\bibfnamefont {J.}~\bibnamefont {Zupan}},\
  }\href@noop {} {\  (\bibinfo {year} {2024})},\ \Eprint
  {http://arxiv.org/abs/2406.01705} {arXiv:2406.01705 [hep-ph]} \BibitemShut
  {NoStop}%
\bibitem [{\citenamefont {Di~Mauro}\ \emph {et~al.}(2016)\citenamefont
  {Di~Mauro}, \citenamefont {Donato}, \citenamefont {Fornengo},\ and\
  \citenamefont {Vittino}}]{DiMauro:2015jxa}%
  \BibitemOpen
  \bibfield  {author} {\bibinfo {author} {\bibfnamefont {M.}~\bibnamefont
  {Di~Mauro}}, \bibinfo {author} {\bibfnamefont {F.}~\bibnamefont {Donato}},
  \bibinfo {author} {\bibfnamefont {N.}~\bibnamefont {Fornengo}}, \ and\
  \bibinfo {author} {\bibfnamefont {A.}~\bibnamefont {Vittino}},\ }\href
  {\doibase 10.1088/1475-7516/2016/05/031} {\bibfield  {journal} {\bibinfo
  {journal} {JCAP}\ }\textbf {\bibinfo {volume} {05}},\ \bibinfo {pages} {031}
  (\bibinfo {year} {2016})},\ \Eprint {http://arxiv.org/abs/1507.07001}
  {arXiv:1507.07001 [astro-ph.HE]} \BibitemShut {NoStop}%
\bibitem [{\citenamefont {Di~Mauro}\ and\ \citenamefont
  {Donato}(2015)}]{DiMauro:2015tfa}%
  \BibitemOpen
  \bibfield  {author} {\bibinfo {author} {\bibfnamefont {M.}~\bibnamefont
  {Di~Mauro}}\ and\ \bibinfo {author} {\bibfnamefont {F.}~\bibnamefont
  {Donato}},\ }\href {\doibase 10.1103/PhysRevD.91.123001} {\bibfield
  {journal} {\bibinfo  {journal} {Phys. Rev. D}\ }\textbf {\bibinfo {volume}
  {91}},\ \bibinfo {pages} {123001} (\bibinfo {year} {2015})},\ \Eprint
  {http://arxiv.org/abs/1501.05316} {arXiv:1501.05316 [astro-ph.HE]}
  \BibitemShut {NoStop}%
\bibitem [{\citenamefont {G\'enolini}\ \emph {et~al.}(2021)\citenamefont
  {G\'enolini}, \citenamefont {Boudaud}, \citenamefont {Cirelli}, \citenamefont
  {Derome}, \citenamefont {Lavalle}, \citenamefont {Maurin}, \citenamefont
  {Salati},\ and\ \citenamefont {Weinrich}}]{Genolini:2021doh}%
  \BibitemOpen
  \bibfield  {author} {\bibinfo {author} {\bibfnamefont {Y.}~\bibnamefont
  {G\'enolini}}, \bibinfo {author} {\bibfnamefont {M.}~\bibnamefont {Boudaud}},
  \bibinfo {author} {\bibfnamefont {M.}~\bibnamefont {Cirelli}}, \bibinfo
  {author} {\bibfnamefont {L.}~\bibnamefont {Derome}}, \bibinfo {author}
  {\bibfnamefont {J.}~\bibnamefont {Lavalle}}, \bibinfo {author} {\bibfnamefont
  {D.}~\bibnamefont {Maurin}}, \bibinfo {author} {\bibfnamefont
  {P.}~\bibnamefont {Salati}}, \ and\ \bibinfo {author} {\bibfnamefont
  {N.}~\bibnamefont {Weinrich}},\ }\href {\doibase 10.1103/PhysRevD.104.083005}
  {\bibfield  {journal} {\bibinfo  {journal} {Phys. Rev. D}\ }\textbf {\bibinfo
  {volume} {104}},\ \bibinfo {pages} {083005} (\bibinfo {year} {2021})},\
  \Eprint {http://arxiv.org/abs/2103.04108} {arXiv:2103.04108 [astro-ph.HE]}
  \BibitemShut {NoStop}%
\bibitem [{\citenamefont {Di~Mauro}\ and\ \citenamefont
  {Winkler}(2021)}]{DiMauro:2021qcf}%
  \BibitemOpen
  \bibfield  {author} {\bibinfo {author} {\bibfnamefont {M.}~\bibnamefont
  {Di~Mauro}}\ and\ \bibinfo {author} {\bibfnamefont {M.~W.}\ \bibnamefont
  {Winkler}},\ }\href {\doibase 10.1103/PhysRevD.103.123005} {\bibfield
  {journal} {\bibinfo  {journal} {Phys. Rev. D}\ }\textbf {\bibinfo {volume}
  {103}},\ \bibinfo {pages} {123005} (\bibinfo {year} {2021})},\ \Eprint
  {http://arxiv.org/abs/2101.11027} {arXiv:2101.11027 [astro-ph.HE]}
  \BibitemShut {NoStop}%
\bibitem [{\citenamefont {McDaniel}\ \emph {et~al.}(2024)\citenamefont
  {McDaniel}, \citenamefont {Ajello}, \citenamefont {Karwin}, \citenamefont
  {Di~Mauro}, \citenamefont {Drlica-Wagner},\ and\ \citenamefont
  {S\'anchez-Conde}}]{McDaniel:2023bju}%
  \BibitemOpen
  \bibfield  {author} {\bibinfo {author} {\bibfnamefont {A.}~\bibnamefont
  {McDaniel}}, \bibinfo {author} {\bibfnamefont {M.}~\bibnamefont {Ajello}},
  \bibinfo {author} {\bibfnamefont {C.~M.}\ \bibnamefont {Karwin}}, \bibinfo
  {author} {\bibfnamefont {M.}~\bibnamefont {Di~Mauro}}, \bibinfo {author}
  {\bibfnamefont {A.}~\bibnamefont {Drlica-Wagner}}, \ and\ \bibinfo {author}
  {\bibfnamefont {M.~A.}\ \bibnamefont {S\'anchez-Conde}},\ }\href {\doibase
  10.1103/PhysRevD.109.063024} {\bibfield  {journal} {\bibinfo  {journal}
  {Phys. Rev. D}\ }\textbf {\bibinfo {volume} {109}},\ \bibinfo {pages}
  {063024} (\bibinfo {year} {2024})},\ \Eprint
  {http://arxiv.org/abs/2311.04982} {arXiv:2311.04982 [astro-ph.HE]}
  \BibitemShut {NoStop}%
\bibitem [{\citenamefont {Balan}\ \emph {et~al.}(2023)\citenamefont {Balan},
  \citenamefont {Kahlhoefer}, \citenamefont {Korsmeier}, \citenamefont
  {Manconi},\ and\ \citenamefont {Nippel}}]{Balan:2023lwg}%
  \BibitemOpen
  \bibfield  {author} {\bibinfo {author} {\bibfnamefont {S.}~\bibnamefont
  {Balan}}, \bibinfo {author} {\bibfnamefont {F.}~\bibnamefont {Kahlhoefer}},
  \bibinfo {author} {\bibfnamefont {M.}~\bibnamefont {Korsmeier}}, \bibinfo
  {author} {\bibfnamefont {S.}~\bibnamefont {Manconi}}, \ and\ \bibinfo
  {author} {\bibfnamefont {K.}~\bibnamefont {Nippel}},\ }\href {\doibase
  10.1088/1475-7516/2023/08/052} {\bibfield  {journal} {\bibinfo  {journal}
  {JCAP}\ }\textbf {\bibinfo {volume} {08}},\ \bibinfo {pages} {052} (\bibinfo
  {year} {2023})},\ \Eprint {http://arxiv.org/abs/2303.07362} {arXiv:2303.07362
  [hep-ph]} \BibitemShut {NoStop}%
\bibitem [{\citenamefont {Donato}\ \emph {et~al.}(2000)\citenamefont {Donato},
  \citenamefont {Fornengo},\ and\ \citenamefont {Salati}}]{Donato:1999gy}%
  \BibitemOpen
  \bibfield  {author} {\bibinfo {author} {\bibfnamefont {F.}~\bibnamefont
  {Donato}}, \bibinfo {author} {\bibfnamefont {N.}~\bibnamefont {Fornengo}}, \
  and\ \bibinfo {author} {\bibfnamefont {P.}~\bibnamefont {Salati}},\ }\href
  {\doibase 10.1103/PhysRevD.62.043003} {\bibfield  {journal} {\bibinfo
  {journal} {Phys. Rev.}\ }\textbf {\bibinfo {volume} {D62}},\ \bibinfo {pages}
  {043003} (\bibinfo {year} {2000})},\ \Eprint
  {http://arxiv.org/abs/hep-ph/9904481} {arXiv:hep-ph/9904481 [hep-ph]}
  \BibitemShut {NoStop}%
%%CITATION = HEP-PH/9904481;%%
\bibitem [{\citenamefont {Cirelli}\ \emph {et~al.}(2014)\citenamefont
  {Cirelli}, \citenamefont {Fornengo}, \citenamefont {Taoso},\ and\
  \citenamefont {Vittino}}]{Cirelli:2014qia}%
  \BibitemOpen
  \bibfield  {author} {\bibinfo {author} {\bibfnamefont {M.}~\bibnamefont
  {Cirelli}}, \bibinfo {author} {\bibfnamefont {N.}~\bibnamefont {Fornengo}},
  \bibinfo {author} {\bibfnamefont {M.}~\bibnamefont {Taoso}}, \ and\ \bibinfo
  {author} {\bibfnamefont {A.}~\bibnamefont {Vittino}},\ }\href {\doibase
  10.1007/JHEP08(2014)009} {\bibfield  {journal} {\bibinfo  {journal} {JHEP}\
  }\textbf {\bibinfo {volume} {08}},\ \bibinfo {pages} {009} (\bibinfo {year}
  {2014})},\ \Eprint {http://arxiv.org/abs/1401.4017} {arXiv:1401.4017
  [hep-ph]} \BibitemShut {NoStop}%
\bibitem [{\citenamefont {Carlson}\ \emph {et~al.}(2014)\citenamefont
  {Carlson}, \citenamefont {Coogan}, \citenamefont {Linden}, \citenamefont
  {Profumo}, \citenamefont {Ibarra},\ and\ \citenamefont
  {Wild}}]{Carlson:2014ssa}%
  \BibitemOpen
  \bibfield  {author} {\bibinfo {author} {\bibfnamefont {E.}~\bibnamefont
  {Carlson}}, \bibinfo {author} {\bibfnamefont {A.}~\bibnamefont {Coogan}},
  \bibinfo {author} {\bibfnamefont {T.}~\bibnamefont {Linden}}, \bibinfo
  {author} {\bibfnamefont {S.}~\bibnamefont {Profumo}}, \bibinfo {author}
  {\bibfnamefont {A.}~\bibnamefont {Ibarra}}, \ and\ \bibinfo {author}
  {\bibfnamefont {S.}~\bibnamefont {Wild}},\ }\href {\doibase
  10.1103/PhysRevD.89.076005} {\bibfield  {journal} {\bibinfo  {journal} {Phys.
  Rev. D}\ }\textbf {\bibinfo {volume} {89}},\ \bibinfo {pages} {076005}
  (\bibinfo {year} {2014})},\ \Eprint {http://arxiv.org/abs/1401.2461}
  {arXiv:1401.2461 [hep-ph]} \BibitemShut {NoStop}%
\bibitem [{\citenamefont {Ibarra}\ and\ \citenamefont
  {Wild}(2013)}]{Ibarra:2012cc}%
  \BibitemOpen
  \bibfield  {author} {\bibinfo {author} {\bibfnamefont {A.}~\bibnamefont
  {Ibarra}}\ and\ \bibinfo {author} {\bibfnamefont {S.}~\bibnamefont {Wild}},\
  }\href {\doibase 10.1088/1475-7516/2013/02/021} {\bibfield  {journal}
  {\bibinfo  {journal} {JCAP}\ }\textbf {\bibinfo {volume} {02}},\ \bibinfo
  {pages} {021} (\bibinfo {year} {2013})},\ \Eprint
  {http://arxiv.org/abs/1209.5539} {arXiv:1209.5539 [hep-ph]} \BibitemShut
  {NoStop}%
\bibitem [{\citenamefont {Fornengo}\ \emph {et~al.}(2013)\citenamefont
  {Fornengo}, \citenamefont {Maccione},\ and\ \citenamefont
  {Vittino}}]{Fornengo:2013osa}%
  \BibitemOpen
  \bibfield  {author} {\bibinfo {author} {\bibfnamefont {N.}~\bibnamefont
  {Fornengo}}, \bibinfo {author} {\bibfnamefont {L.}~\bibnamefont {Maccione}},
  \ and\ \bibinfo {author} {\bibfnamefont {A.}~\bibnamefont {Vittino}},\ }\href
  {\doibase 10.1088/1475-7516/2013/09/031} {\bibfield  {journal} {\bibinfo
  {journal} {JCAP}\ }\textbf {\bibinfo {volume} {09}},\ \bibinfo {pages} {031}
  (\bibinfo {year} {2013})},\ \Eprint {http://arxiv.org/abs/1306.4171}
  {arXiv:1306.4171 [hep-ph]} \BibitemShut {NoStop}%
\bibitem [{\citenamefont {Herms}\ \emph {et~al.}(2017)\citenamefont {Herms},
  \citenamefont {Ibarra}, \citenamefont {Vittino},\ and\ \citenamefont
  {Wild}}]{Herms:2016vop}%
  \BibitemOpen
  \bibfield  {author} {\bibinfo {author} {\bibfnamefont {J.}~\bibnamefont
  {Herms}}, \bibinfo {author} {\bibfnamefont {A.}~\bibnamefont {Ibarra}},
  \bibinfo {author} {\bibfnamefont {A.}~\bibnamefont {Vittino}}, \ and\
  \bibinfo {author} {\bibfnamefont {S.}~\bibnamefont {Wild}},\ }\href {\doibase
  10.1088/1475-7516/2017/02/018} {\bibfield  {journal} {\bibinfo  {journal}
  {JCAP}\ }\textbf {\bibinfo {volume} {02}},\ \bibinfo {pages} {018} (\bibinfo
  {year} {2017})},\ \Eprint {http://arxiv.org/abs/1610.00699} {arXiv:1610.00699
  [astro-ph.HE]} \BibitemShut {NoStop}%
\bibitem [{\citenamefont {Korsmeier}\ \emph {et~al.}(2018)\citenamefont
  {Korsmeier}, \citenamefont {Donato},\ and\ \citenamefont
  {Fornengo}}]{Korsmeier:2017xzj}%
  \BibitemOpen
  \bibfield  {author} {\bibinfo {author} {\bibfnamefont {M.}~\bibnamefont
  {Korsmeier}}, \bibinfo {author} {\bibfnamefont {F.}~\bibnamefont {Donato}}, \
  and\ \bibinfo {author} {\bibfnamefont {N.}~\bibnamefont {Fornengo}},\ }\href
  {\doibase 10.1103/PhysRevD.97.103011} {\bibfield  {journal} {\bibinfo
  {journal} {Phys. Rev. D}\ }\textbf {\bibinfo {volume} {97}},\ \bibinfo
  {pages} {103011} (\bibinfo {year} {2018})},\ \Eprint
  {http://arxiv.org/abs/1711.08465} {arXiv:1711.08465 [astro-ph.HE]}
  \BibitemShut {NoStop}%
\bibitem [{\citenamefont {von Doetinchem}\ \emph {et~al.}(2016)\citenamefont
  {von Doetinchem} \emph {et~al.}}]{vonDoetinchem:2015yva}%
  \BibitemOpen
  \bibfield  {author} {\bibinfo {author} {\bibfnamefont {P.}~\bibnamefont {von
  Doetinchem}} \emph {et~al.},\ }\href {\doibase 10.22323/1.236.1218}
  {\bibfield  {journal} {\bibinfo  {journal} {PoS}\ }\textbf {\bibinfo {volume}
  {ICRC2015}},\ \bibinfo {pages} {1218} (\bibinfo {year} {2016})},\ \Eprint
  {http://arxiv.org/abs/1507.02712} {arXiv:1507.02712 [hep-ph]} \BibitemShut
  {NoStop}%
\bibitem [{\citenamefont {{Choutko}}\ and\ \citenamefont
  {{Giovacchini}}(2008)}]{2008ICRC....4..765C}%
  \BibitemOpen
  \bibfield  {author} {\bibinfo {author} {\bibfnamefont {V.}~\bibnamefont
  {{Choutko}}}\ and\ \bibinfo {author} {\bibfnamefont {F.}~\bibnamefont
  {{Giovacchini}}},\ }in\ \href@noop {} {\emph {\bibinfo {booktitle}
  {International Cosmic Ray Conference}}},\ \bibinfo {series} {International
  Cosmic Ray Conference}, Vol.~\bibinfo {volume} {4}\ (\bibinfo {year} {2008})\
  pp.\ \bibinfo {pages} {765--768}\BibitemShut {NoStop}%
\bibitem [{\citenamefont {Aramaki}\ \emph {et~al.}(2016)\citenamefont
  {Aramaki}, \citenamefont {Hailey}, \citenamefont {Boggs}, \citenamefont {von
  Doetinchem}, \citenamefont {Fuke}, \citenamefont {Mognet}, \citenamefont
  {Ong}, \citenamefont {Perez},\ and\ \citenamefont
  {Zweerink}}]{Aramaki:2015laa}%
  \BibitemOpen
  \bibfield  {author} {\bibinfo {author} {\bibfnamefont {T.}~\bibnamefont
  {Aramaki}}, \bibinfo {author} {\bibfnamefont {C.~J.}\ \bibnamefont {Hailey}},
  \bibinfo {author} {\bibfnamefont {S.~E.}\ \bibnamefont {Boggs}}, \bibinfo
  {author} {\bibfnamefont {P.}~\bibnamefont {von Doetinchem}}, \bibinfo
  {author} {\bibfnamefont {H.}~\bibnamefont {Fuke}}, \bibinfo {author}
  {\bibfnamefont {S.~I.}\ \bibnamefont {Mognet}}, \bibinfo {author}
  {\bibfnamefont {R.~A.}\ \bibnamefont {Ong}}, \bibinfo {author} {\bibfnamefont
  {K.}~\bibnamefont {Perez}}, \ and\ \bibinfo {author} {\bibfnamefont
  {J.}~\bibnamefont {Zweerink}} (\bibinfo {collaboration} {GAPS}),\ }\href
  {\doibase 10.1016/j.astropartphys.2015.09.001} {\bibfield  {journal}
  {\bibinfo  {journal} {Astropart. Phys.}\ }\textbf {\bibinfo {volume} {74}},\
  \bibinfo {pages} {6} (\bibinfo {year} {2016})},\ \Eprint
  {http://arxiv.org/abs/1506.02513} {arXiv:1506.02513 [astro-ph.HE]}
  \BibitemShut {NoStop}%
\bibitem [{\citenamefont {Kapusta}(1980)}]{Kapusta:1980}%
  \BibitemOpen
  \bibfield  {author} {\bibinfo {author} {\bibfnamefont {J.~I.}\ \bibnamefont
  {Kapusta}},\ }\href {\doibase 10.1103/PhysRevC.21.1301} {\bibfield  {journal}
  {\bibinfo  {journal} {Phys. Rev.}\ }\textbf {\bibinfo {volume} {C21}},\
  \bibinfo {pages} {1301} (\bibinfo {year} {1980})}\BibitemShut {NoStop}%
%%CITATION = PHRVA,C21,1301;%%
\bibitem [{\citenamefont {Butler}\ and\ \citenamefont
  {Pearson}(1963)}]{Butler:1963}%
  \BibitemOpen
  \bibfield  {author} {\bibinfo {author} {\bibfnamefont {S.~T.}\ \bibnamefont
  {Butler}}\ and\ \bibinfo {author} {\bibfnamefont {C.~A.}\ \bibnamefont
  {Pearson}},\ }\href {\doibase 10.1103/PhysRev.129.836} {\bibfield  {journal}
  {\bibinfo  {journal} {Phys. Rev.}\ }\textbf {\bibinfo {volume} {129}},\
  \bibinfo {pages} {836} (\bibinfo {year} {1963})}\BibitemShut {NoStop}%
%%CITATION = PHRVA,129,836;%%
\bibitem [{\citenamefont {Scheibl}\ and\ \citenamefont
  {Heinz}(1999)}]{Scheibl:1998tk}%
  \BibitemOpen
  \bibfield  {author} {\bibinfo {author} {\bibfnamefont {R.}~\bibnamefont
  {Scheibl}}\ and\ \bibinfo {author} {\bibfnamefont {U.~W.}\ \bibnamefont
  {Heinz}},\ }\href {\doibase 10.1103/PhysRevC.59.1585} {\bibfield  {journal}
  {\bibinfo  {journal} {Phys. Rev. C}\ }\textbf {\bibinfo {volume} {59}},\
  \bibinfo {pages} {1585} (\bibinfo {year} {1999})},\ \Eprint
  {http://arxiv.org/abs/nucl-th/9809092} {arXiv:nucl-th/9809092} \BibitemShut
  {NoStop}%
\bibitem [{\citenamefont {Kachelrie\ss{}}\ \emph
  {et~al.}(2020{\natexlab{a}})\citenamefont {Kachelrie\ss{}}, \citenamefont
  {Ostapchenko},\ and\ \citenamefont {Tjemsland}}]{Kachelriess:2020uoh}%
  \BibitemOpen
  \bibfield  {author} {\bibinfo {author} {\bibfnamefont {M.}~\bibnamefont
  {Kachelrie\ss{}}}, \bibinfo {author} {\bibfnamefont {S.}~\bibnamefont
  {Ostapchenko}}, \ and\ \bibinfo {author} {\bibfnamefont {J.}~\bibnamefont
  {Tjemsland}},\ }\href {\doibase 10.1088/1475-7516/2020/08/048} {\bibfield
  {journal} {\bibinfo  {journal} {JCAP}\ }\textbf {\bibinfo {volume} {08}},\
  \bibinfo {pages} {048} (\bibinfo {year} {2020}{\natexlab{a}})},\ \Eprint
  {http://arxiv.org/abs/2002.10481} {arXiv:2002.10481 [hep-ph]} \BibitemShut
  {NoStop}%
\bibitem [{\citenamefont {Schael}\ \emph
  {et~al.}(2006{\natexlab{a}})\citenamefont {Schael}, \citenamefont {Barate},
  \citenamefont {Brunelière} \emph {et~al.}}]{2006192}%
  \BibitemOpen
  \bibfield  {author} {\bibinfo {author} {\bibfnamefont {S.}~\bibnamefont
  {Schael}}, \bibinfo {author} {\bibfnamefont {R.}~\bibnamefont {Barate}},
  \bibinfo {author} {\bibfnamefont {R.}~\bibnamefont {Brunelière}},  \emph
  {et~al.},\ }\href {\doibase https://doi.org/10.1016/j.physletb.2006.06.043}
  {\bibfield  {journal} {\bibinfo  {journal} {Physics Letters B}\ }\textbf
  {\bibinfo {volume} {639}},\ \bibinfo {pages} {192} (\bibinfo {year}
  {2006}{\natexlab{a}})}\BibitemShut {NoStop}%
\bibitem [{\citenamefont {{Akers}}\ \emph {et~al.}(1995)\citenamefont
  {{Akers}}, \citenamefont {{Alexander}}, \citenamefont {{Allison}} \emph
  {et~al.}}]{1995ZPhyC..67..203A}%
  \BibitemOpen
  \bibfield  {author} {\bibinfo {author} {\bibfnamefont {R.}~\bibnamefont
  {{Akers}}}, \bibinfo {author} {\bibfnamefont {G.}~\bibnamefont
  {{Alexander}}}, \bibinfo {author} {\bibfnamefont {J.}~\bibnamefont
  {{Allison}}},  \emph {et~al.},\ }\href {\doibase 10.1007/BF01571281}
  {\bibfield  {journal} {\bibinfo  {journal} {Zeitschrift fur Physik C
  Particles and Fields}\ }\textbf {\bibinfo {volume} {67}},\ \bibinfo {pages}
  {203} (\bibinfo {year} {1995})}\BibitemShut {NoStop}%
\bibitem [{\citenamefont {Adam}\ \emph {et~al.}(2016)\citenamefont {Adam} \emph
  {et~al.}}]{ALICE:2015wav}%
  \BibitemOpen
  \bibfield  {author} {\bibinfo {author} {\bibfnamefont {J.}~\bibnamefont
  {Adam}} \emph {et~al.} (\bibinfo {collaboration} {ALICE}),\ }\href {\doibase
  10.1103/PhysRevC.93.024917} {\bibfield  {journal} {\bibinfo  {journal} {Phys.
  Rev. C}\ }\textbf {\bibinfo {volume} {93}},\ \bibinfo {pages} {024917}
  (\bibinfo {year} {2016})},\ \Eprint {http://arxiv.org/abs/1506.08951}
  {arXiv:1506.08951 [nucl-ex]} \BibitemShut {NoStop}%
\bibitem [{\citenamefont {Acharya}\ \emph {et~al.}(2018)\citenamefont {Acharya}
  \emph {et~al.}}]{ALICE:2017xrp}%
  \BibitemOpen
  \bibfield  {author} {\bibinfo {author} {\bibfnamefont {S.}~\bibnamefont
  {Acharya}} \emph {et~al.} (\bibinfo {collaboration} {ALICE}),\ }\href
  {\doibase 10.1103/PhysRevC.97.024615} {\bibfield  {journal} {\bibinfo
  {journal} {Phys. Rev. C}\ }\textbf {\bibinfo {volume} {97}},\ \bibinfo
  {pages} {024615} (\bibinfo {year} {2018})},\ \Eprint
  {http://arxiv.org/abs/1709.08522} {arXiv:1709.08522 [nucl-ex]} \BibitemShut
  {NoStop}%
\bibitem [{\citenamefont {Acharya}\ \emph {et~al.}(2019)\citenamefont {Acharya}
  \emph {et~al.}}]{ALICE:2019dgz}%
  \BibitemOpen
  \bibfield  {author} {\bibinfo {author} {\bibfnamefont {S.}~\bibnamefont
  {Acharya}} \emph {et~al.} (\bibinfo {collaboration} {ALICE}),\ }\href
  {\doibase 10.1016/j.physletb.2019.05.028} {\bibfield  {journal} {\bibinfo
  {journal} {Phys. Lett. B}\ }\textbf {\bibinfo {volume} {794}},\ \bibinfo
  {pages} {50} (\bibinfo {year} {2019})},\ \Eprint
  {http://arxiv.org/abs/1902.09290} {arXiv:1902.09290 [nucl-ex]} \BibitemShut
  {NoStop}%
\bibitem [{\citenamefont {Acharya}\ \emph {et~al.}(2020)\citenamefont {Acharya}
  \emph {et~al.}}]{ALICE:2020foi}%
  \BibitemOpen
  \bibfield  {author} {\bibinfo {author} {\bibfnamefont {S.}~\bibnamefont
  {Acharya}} \emph {et~al.} (\bibinfo {collaboration} {ALICE}),\ }\href
  {\doibase 10.1140/epjc/s10052-020-8256-4} {\bibfield  {journal} {\bibinfo
  {journal} {Eur. Phys. J. C}\ }\textbf {\bibinfo {volume} {80}},\ \bibinfo
  {pages} {889} (\bibinfo {year} {2020})},\ \Eprint
  {http://arxiv.org/abs/2003.03184} {arXiv:2003.03184 [nucl-ex]} \BibitemShut
  {NoStop}%
\bibitem [{\citenamefont {Acharya}\ \emph
  {et~al.}(2022{\natexlab{a}})\citenamefont {Acharya} \emph
  {et~al.}}]{ALICE:2021mfm}%
  \BibitemOpen
  \bibfield  {author} {\bibinfo {author} {\bibfnamefont {S.}~\bibnamefont
  {Acharya}} \emph {et~al.} (\bibinfo {collaboration} {ALICE}),\ }\href
  {\doibase 10.1007/JHEP01(2022)106} {\bibfield  {journal} {\bibinfo  {journal}
  {JHEP}\ }\textbf {\bibinfo {volume} {01}},\ \bibinfo {pages} {106} (\bibinfo
  {year} {2022}{\natexlab{a}})},\ \Eprint {http://arxiv.org/abs/2109.13026}
  {arXiv:2109.13026 [nucl-ex]} \BibitemShut {NoStop}%
\bibitem [{\citenamefont {Acharya}\ \emph
  {et~al.}(2022{\natexlab{b}})\citenamefont {Acharya} \emph
  {et~al.}}]{ALICE:2021ovi}%
  \BibitemOpen
  \bibfield  {author} {\bibinfo {author} {\bibfnamefont {S.}~\bibnamefont
  {Acharya}} \emph {et~al.} (\bibinfo {collaboration} {ALICE}),\ }\href
  {\doibase 10.1140/epjc/s10052-022-10241-z} {\bibfield  {journal} {\bibinfo
  {journal} {Eur. Phys. J. C}\ }\textbf {\bibinfo {volume} {82}},\ \bibinfo
  {pages} {289} (\bibinfo {year} {2022}{\natexlab{b}})},\ \Eprint
  {http://arxiv.org/abs/2112.00610} {arXiv:2112.00610 [nucl-ex]} \BibitemShut
  {NoStop}%
\bibitem [{\citenamefont {Bellini}\ and\ \citenamefont
  {Kalweit}(2019)}]{Bellini:2018epz}%
  \BibitemOpen
  \bibfield  {author} {\bibinfo {author} {\bibfnamefont {F.}~\bibnamefont
  {Bellini}}\ and\ \bibinfo {author} {\bibfnamefont {A.~P.}\ \bibnamefont
  {Kalweit}},\ }\href {\doibase 10.1103/PhysRevC.99.054905} {\bibfield
  {journal} {\bibinfo  {journal} {Phys. Rev. C}\ }\textbf {\bibinfo {volume}
  {99}},\ \bibinfo {pages} {054905} (\bibinfo {year} {2019})},\ \Eprint
  {http://arxiv.org/abs/1807.05894} {arXiv:1807.05894 [hep-ph]} \BibitemShut
  {NoStop}%
\bibitem [{\citenamefont {Blum}\ \emph {et~al.}(2017)\citenamefont {Blum},
  \citenamefont {Ng}, \citenamefont {Sato},\ and\ \citenamefont
  {Takimoto}}]{Blum:2017qnn}%
  \BibitemOpen
  \bibfield  {author} {\bibinfo {author} {\bibfnamefont {K.}~\bibnamefont
  {Blum}}, \bibinfo {author} {\bibfnamefont {K.~C.~Y.}\ \bibnamefont {Ng}},
  \bibinfo {author} {\bibfnamefont {R.}~\bibnamefont {Sato}}, \ and\ \bibinfo
  {author} {\bibfnamefont {M.}~\bibnamefont {Takimoto}},\ }\href {\doibase
  10.1103/PhysRevD.96.103021} {\bibfield  {journal} {\bibinfo  {journal} {Phys.
  Rev. D}\ }\textbf {\bibinfo {volume} {96}},\ \bibinfo {pages} {103021}
  (\bibinfo {year} {2017})},\ \Eprint {http://arxiv.org/abs/1704.05431}
  {arXiv:1704.05431 [astro-ph.HE]} \BibitemShut {NoStop}%
\bibitem [{\citenamefont {Bellini}\ \emph {et~al.}(2021)\citenamefont
  {Bellini}, \citenamefont {Blum}, \citenamefont {Kalweit},\ and\ \citenamefont
  {Puccio}}]{Bellini:2020cbj}%
  \BibitemOpen
  \bibfield  {author} {\bibinfo {author} {\bibfnamefont {F.}~\bibnamefont
  {Bellini}}, \bibinfo {author} {\bibfnamefont {K.}~\bibnamefont {Blum}},
  \bibinfo {author} {\bibfnamefont {A.~P.}\ \bibnamefont {Kalweit}}, \ and\
  \bibinfo {author} {\bibfnamefont {M.}~\bibnamefont {Puccio}},\ }\href
  {\doibase 10.1103/PhysRevC.103.014907} {\bibfield  {journal} {\bibinfo
  {journal} {Phys. Rev. C}\ }\textbf {\bibinfo {volume} {103}},\ \bibinfo
  {pages} {014907} (\bibinfo {year} {2021})},\ \Eprint
  {http://arxiv.org/abs/2007.01750} {arXiv:2007.01750 [nucl-th]} \BibitemShut
  {NoStop}%
\bibitem [{\citenamefont {Kachelrie\ss{}}\ \emph
  {et~al.}(2020{\natexlab{b}})\citenamefont {Kachelrie\ss{}}, \citenamefont
  {Ostapchenko},\ and\ \citenamefont {Tjemsland}}]{Kachelriess:2019taq}%
  \BibitemOpen
  \bibfield  {author} {\bibinfo {author} {\bibfnamefont {M.}~\bibnamefont
  {Kachelrie\ss{}}}, \bibinfo {author} {\bibfnamefont {S.}~\bibnamefont
  {Ostapchenko}}, \ and\ \bibinfo {author} {\bibfnamefont {J.}~\bibnamefont
  {Tjemsland}},\ }\href {\doibase 10.1140/epja/s10050-019-00007-9} {\bibfield
  {journal} {\bibinfo  {journal} {Eur. Phys. J. A}\ }\textbf {\bibinfo {volume}
  {56}},\ \bibinfo {pages} {4} (\bibinfo {year} {2020}{\natexlab{b}})},\
  \Eprint {http://arxiv.org/abs/1905.01192} {arXiv:1905.01192 [hep-ph]}
  \BibitemShut {NoStop}%
\bibitem [{\citenamefont {Mahlein}\ \emph {et~al.}(2023)\citenamefont
  {Mahlein}, \citenamefont {Barioglio}, \citenamefont {Bellini}, \citenamefont
  {Fabbietti}, \citenamefont {Pinto}, \citenamefont {Singh},\ and\
  \citenamefont {Tripathy}}]{Mahlein:2023fmx}%
  \BibitemOpen
  \bibfield  {author} {\bibinfo {author} {\bibfnamefont {M.}~\bibnamefont
  {Mahlein}}, \bibinfo {author} {\bibfnamefont {L.}~\bibnamefont {Barioglio}},
  \bibinfo {author} {\bibfnamefont {F.}~\bibnamefont {Bellini}}, \bibinfo
  {author} {\bibfnamefont {L.}~\bibnamefont {Fabbietti}}, \bibinfo {author}
  {\bibfnamefont {C.}~\bibnamefont {Pinto}}, \bibinfo {author} {\bibfnamefont
  {B.}~\bibnamefont {Singh}}, \ and\ \bibinfo {author} {\bibfnamefont
  {S.}~\bibnamefont {Tripathy}},\ }\href {\doibase
  10.1140/epjc/s10052-023-11972-3} {\bibfield  {journal} {\bibinfo  {journal}
  {Eur. Phys. J. C}\ }\textbf {\bibinfo {volume} {83}},\ \bibinfo {pages} {804}
  (\bibinfo {year} {2023})},\ \Eprint {http://arxiv.org/abs/2302.12696}
  {arXiv:2302.12696 [hep-ex]} \BibitemShut {NoStop}%
\bibitem [{\citenamefont {Wiringa}\ \emph {et~al.}(1995)\citenamefont
  {Wiringa}, \citenamefont {Stoks},\ and\ \citenamefont
  {Schiavilla}}]{Wiringa:1994wb}%
  \BibitemOpen
  \bibfield  {author} {\bibinfo {author} {\bibfnamefont {R.~B.}\ \bibnamefont
  {Wiringa}}, \bibinfo {author} {\bibfnamefont {V.~G.~J.}\ \bibnamefont
  {Stoks}}, \ and\ \bibinfo {author} {\bibfnamefont {R.}~\bibnamefont
  {Schiavilla}},\ }\href {\doibase 10.1103/PhysRevC.51.38} {\bibfield
  {journal} {\bibinfo  {journal} {Phys. Rev. C}\ }\textbf {\bibinfo {volume}
  {51}},\ \bibinfo {pages} {38} (\bibinfo {year} {1995})},\ \Eprint
  {http://arxiv.org/abs/nucl-th/9408016} {arXiv:nucl-th/9408016} \BibitemShut
  {NoStop}%
\bibitem [{\citenamefont {Schael}\ \emph
  {et~al.}(2006{\natexlab{b}})\citenamefont {Schael} \emph
  {et~al.}}]{ALEPH:2006qoi}%
  \BibitemOpen
  \bibfield  {author} {\bibinfo {author} {\bibfnamefont {S.}~\bibnamefont
  {Schael}} \emph {et~al.} (\bibinfo {collaboration} {ALEPH}),\ }\href
  {\doibase 10.1016/j.physletb.2006.06.043} {\bibfield  {journal} {\bibinfo
  {journal} {Phys. Lett. B}\ }\textbf {\bibinfo {volume} {639}},\ \bibinfo
  {pages} {192} (\bibinfo {year} {2006}{\natexlab{b}})},\ \Eprint
  {http://arxiv.org/abs/hep-ex/0604023} {arXiv:hep-ex/0604023} \BibitemShut
  {NoStop}%
\bibitem [{\citenamefont {Bierlich}\ \emph {et~al.}(2022)\citenamefont
  {Bierlich} \emph {et~al.}}]{Bierlich:2022pfr}%
  \BibitemOpen
  \bibfield  {author} {\bibinfo {author} {\bibfnamefont {C.}~\bibnamefont
  {Bierlich}} \emph {et~al.},\ }\href {\doibase 10.21468/SciPostPhysCodeb.8} {\
   (\bibinfo {year} {2022}),\ 10.21468/SciPostPhysCodeb.8},\ \Eprint
  {http://arxiv.org/abs/2203.11601} {arXiv:2203.11601 [hep-ph]} \BibitemShut
  {NoStop}%
\bibitem [{\citenamefont {Fischer}\ \emph {et~al.}(2016)\citenamefont
  {Fischer}, \citenamefont {Prestel}, \citenamefont {Ritzmann},\ and\
  \citenamefont {Skands}}]{Fischer:2016vfv}%
  \BibitemOpen
  \bibfield  {author} {\bibinfo {author} {\bibfnamefont {N.}~\bibnamefont
  {Fischer}}, \bibinfo {author} {\bibfnamefont {S.}~\bibnamefont {Prestel}},
  \bibinfo {author} {\bibfnamefont {M.}~\bibnamefont {Ritzmann}}, \ and\
  \bibinfo {author} {\bibfnamefont {P.}~\bibnamefont {Skands}},\ }\href
  {\doibase 10.1140/epjc/s10052-016-4429-6} {\bibfield  {journal} {\bibinfo
  {journal} {Eur. Phys. J. C}\ }\textbf {\bibinfo {volume} {76}},\ \bibinfo
  {pages} {589} (\bibinfo {year} {2016})},\ \Eprint
  {http://arxiv.org/abs/1605.06142} {arXiv:1605.06142 [hep-ph]} \BibitemShut
  {NoStop}%
\bibitem [{\citenamefont {Arina}\ \emph {et~al.}(2024)\citenamefont {Arina},
  \citenamefont {Di~Mauro}, \citenamefont {Fornengo}, \citenamefont {Heisig},
  \citenamefont {Jueid},\ and\ \citenamefont {de~Austri}}]{Arina:2023eic}%
  \BibitemOpen
  \bibfield  {author} {\bibinfo {author} {\bibfnamefont {C.}~\bibnamefont
  {Arina}}, \bibinfo {author} {\bibfnamefont {M.}~\bibnamefont {Di~Mauro}},
  \bibinfo {author} {\bibfnamefont {N.}~\bibnamefont {Fornengo}}, \bibinfo
  {author} {\bibfnamefont {J.}~\bibnamefont {Heisig}}, \bibinfo {author}
  {\bibfnamefont {A.}~\bibnamefont {Jueid}}, \ and\ \bibinfo {author}
  {\bibfnamefont {R.~R.}\ \bibnamefont {de~Austri}},\ }\href {\doibase
  10.1088/1475-7516/2024/03/035} {\bibfield  {journal} {\bibinfo  {journal}
  {JCAP}\ }\textbf {\bibinfo {volume} {03}},\ \bibinfo {pages} {035} (\bibinfo
  {year} {2024})},\ \Eprint {http://arxiv.org/abs/2312.01153} {arXiv:2312.01153
  [astro-ph.HE]} \BibitemShut {NoStop}%
\bibitem [{\citenamefont {Pohl}\ \emph {et~al.}(2016)\citenamefont {Pohl} \emph
  {et~al.}}]{CREMA:2016idx}%
  \BibitemOpen
  \bibfield  {author} {\bibinfo {author} {\bibfnamefont {R.}~\bibnamefont
  {Pohl}} \emph {et~al.} (\bibinfo {collaboration} {CREMA}),\ }\href {\doibase
  10.1126/science.aaf2468} {\bibfield  {journal} {\bibinfo  {journal}
  {Science}\ }\textbf {\bibinfo {volume} {353}},\ \bibinfo {pages} {669}
  (\bibinfo {year} {2016})}\BibitemShut {NoStop}%
\bibitem [{\citenamefont {Cirelli}\ \emph {et~al.}(2011)\citenamefont
  {Cirelli}, \citenamefont {Corcella}, \citenamefont {Hektor}, \citenamefont
  {Hutsi}, \citenamefont {Kadastik}, \citenamefont {Panci}, \citenamefont
  {Raidal}, \citenamefont {Sala},\ and\ \citenamefont
  {Strumia}}]{Cirelli:2010xx}%
  \BibitemOpen
  \bibfield  {author} {\bibinfo {author} {\bibfnamefont {M.}~\bibnamefont
  {Cirelli}}, \bibinfo {author} {\bibfnamefont {G.}~\bibnamefont {Corcella}},
  \bibinfo {author} {\bibfnamefont {A.}~\bibnamefont {Hektor}}, \bibinfo
  {author} {\bibfnamefont {G.}~\bibnamefont {Hutsi}}, \bibinfo {author}
  {\bibfnamefont {M.}~\bibnamefont {Kadastik}}, \bibinfo {author}
  {\bibfnamefont {P.}~\bibnamefont {Panci}}, \bibinfo {author} {\bibfnamefont
  {M.}~\bibnamefont {Raidal}}, \bibinfo {author} {\bibfnamefont
  {F.}~\bibnamefont {Sala}}, \ and\ \bibinfo {author} {\bibfnamefont
  {A.}~\bibnamefont {Strumia}},\ }\href {\doibase
  10.1088/1475-7516/2012/10/E01} {\bibfield  {journal} {\bibinfo  {journal}
  {JCAP}\ }\textbf {\bibinfo {volume} {03}},\ \bibinfo {pages} {051} (\bibinfo
  {year} {2011})},\ \bibinfo {note} {[Erratum: JCAP 10, E01 (2012)]},\ \Eprint
  {http://arxiv.org/abs/1012.4515} {arXiv:1012.4515 [hep-ph]} \BibitemShut
  {NoStop}%
\bibitem [{\citenamefont {Kadastik}\ \emph {et~al.}(2010)\citenamefont
  {Kadastik}, \citenamefont {Raidal},\ and\ \citenamefont
  {Strumia}}]{Kadastik:2009ts}%
  \BibitemOpen
  \bibfield  {author} {\bibinfo {author} {\bibfnamefont {M.}~\bibnamefont
  {Kadastik}}, \bibinfo {author} {\bibfnamefont {M.}~\bibnamefont {Raidal}}, \
  and\ \bibinfo {author} {\bibfnamefont {A.}~\bibnamefont {Strumia}},\ }\href
  {\doibase 10.1016/j.physletb.2009.12.005} {\bibfield  {journal} {\bibinfo
  {journal} {Phys. Lett. B}\ }\textbf {\bibinfo {volume} {683}},\ \bibinfo
  {pages} {248} (\bibinfo {year} {2010})},\ \Eprint
  {http://arxiv.org/abs/0908.1578} {arXiv:0908.1578 [hep-ph]} \BibitemShut
  {NoStop}%
\bibitem [{\citenamefont {Stoks}\ \emph {et~al.}(1993)\citenamefont {Stoks},
  \citenamefont {Klomp}, \citenamefont {Rentmeester},\ and\ \citenamefont
  {de~Swart}}]{PhysRevC.48.792}%
  \BibitemOpen
  \bibfield  {author} {\bibinfo {author} {\bibfnamefont {V.~G.~J.}\
  \bibnamefont {Stoks}}, \bibinfo {author} {\bibfnamefont {R.~A.~M.}\
  \bibnamefont {Klomp}}, \bibinfo {author} {\bibfnamefont {M.~C.~M.}\
  \bibnamefont {Rentmeester}}, \ and\ \bibinfo {author} {\bibfnamefont {J.~J.}\
  \bibnamefont {de~Swart}},\ }\href {\doibase 10.1103/PhysRevC.48.792}
  {\bibfield  {journal} {\bibinfo  {journal} {Phys. Rev. C}\ }\textbf {\bibinfo
  {volume} {48}},\ \bibinfo {pages} {792} (\bibinfo {year} {1993})}\BibitemShut
  {NoStop}%
\bibitem [{\citenamefont {Jueid}\ \emph {et~al.}(2023)\citenamefont {Jueid},
  \citenamefont {Kip}, \citenamefont {de~Austri},\ and\ \citenamefont
  {Skands}}]{Jueid:2022qjg}%
  \BibitemOpen
  \bibfield  {author} {\bibinfo {author} {\bibfnamefont {A.}~\bibnamefont
  {Jueid}}, \bibinfo {author} {\bibfnamefont {J.}~\bibnamefont {Kip}}, \bibinfo
  {author} {\bibfnamefont {R.~R.}\ \bibnamefont {de~Austri}}, \ and\ \bibinfo
  {author} {\bibfnamefont {P.}~\bibnamefont {Skands}},\ }\href {\doibase
  10.1088/1475-7516/2023/04/068} {\bibfield  {journal} {\bibinfo  {journal}
  {JCAP}\ }\textbf {\bibinfo {volume} {04}},\ \bibinfo {pages} {068} (\bibinfo
  {year} {2023})},\ \Eprint {http://arxiv.org/abs/2202.11546} {arXiv:2202.11546
  [hep-ph]} \BibitemShut {NoStop}%
\bibitem [{\citenamefont {Jueid}\ \emph {et~al.}(2024)\citenamefont {Jueid},
  \citenamefont {Kip}, \citenamefont {de~Austri},\ and\ \citenamefont
  {Skands}}]{Jueid:2023vrb}%
  \BibitemOpen
  \bibfield  {author} {\bibinfo {author} {\bibfnamefont {A.}~\bibnamefont
  {Jueid}}, \bibinfo {author} {\bibfnamefont {J.}~\bibnamefont {Kip}}, \bibinfo
  {author} {\bibfnamefont {R.~R.}\ \bibnamefont {de~Austri}}, \ and\ \bibinfo
  {author} {\bibfnamefont {P.}~\bibnamefont {Skands}},\ }\href {\doibase
  10.1007/JHEP02(2024)119} {\bibfield  {journal} {\bibinfo  {journal} {JHEP}\
  }\textbf {\bibinfo {volume} {02}},\ \bibinfo {pages} {119} (\bibinfo {year}
  {2024})},\ \Eprint {http://arxiv.org/abs/2303.11363} {arXiv:2303.11363
  [hep-ph]} \BibitemShut {NoStop}%
\bibitem [{\citenamefont {Horst}\ \emph {et~al.}(2023)\citenamefont {Horst},
  \citenamefont {Barioglio}, \citenamefont {Bellini}, \citenamefont
  {Fabbietti}, \citenamefont {Pinto}, \citenamefont {Singh},\ and\
  \citenamefont {Tripathy}}]{Horst:2023oti}%
  \BibitemOpen
  \bibfield  {author} {\bibinfo {author} {\bibfnamefont {M.}~\bibnamefont
  {Horst}}, \bibinfo {author} {\bibfnamefont {L.}~\bibnamefont {Barioglio}},
  \bibinfo {author} {\bibfnamefont {F.}~\bibnamefont {Bellini}}, \bibinfo
  {author} {\bibfnamefont {L.}~\bibnamefont {Fabbietti}}, \bibinfo {author}
  {\bibfnamefont {C.}~\bibnamefont {Pinto}}, \bibinfo {author} {\bibfnamefont
  {B.}~\bibnamefont {Singh}}, \ and\ \bibinfo {author} {\bibfnamefont
  {S.}~\bibnamefont {Tripathy}},\ }\href@noop {} {\  (\bibinfo {year}
  {2023})},\ \Eprint {http://arxiv.org/abs/2302.12696} {arXiv:2302.12696
  [hep-ex]} \BibitemShut {NoStop}%
\bibitem [{\citenamefont {Winkler}\ and\ \citenamefont
  {Linden}(2021)}]{Winkler:2020ltd}%
  \BibitemOpen
  \bibfield  {author} {\bibinfo {author} {\bibfnamefont {M.~W.}\ \bibnamefont
  {Winkler}}\ and\ \bibinfo {author} {\bibfnamefont {T.}~\bibnamefont
  {Linden}},\ }\href {\doibase 10.1103/PhysRevLett.126.101101} {\bibfield
  {journal} {\bibinfo  {journal} {Phys. Rev. Lett.}\ }\textbf {\bibinfo
  {volume} {126}},\ \bibinfo {pages} {101101} (\bibinfo {year} {2021})},\
  \Eprint {http://arxiv.org/abs/2006.16251} {arXiv:2006.16251 [hep-ph]}
  \BibitemShut {NoStop}%
\bibitem [{\citenamefont {Backovic}\ \emph {et~al.}(2014)\citenamefont
  {Backovic}, \citenamefont {Kong},\ and\ \citenamefont
  {McCaskey}}]{Backovic:2013dpa}%
  \BibitemOpen
  \bibfield  {author} {\bibinfo {author} {\bibfnamefont {M.}~\bibnamefont
  {Backovic}}, \bibinfo {author} {\bibfnamefont {K.}~\bibnamefont {Kong}}, \
  and\ \bibinfo {author} {\bibfnamefont {M.}~\bibnamefont {McCaskey}},\ }\href
  {\doibase 10.1016/j.dark.2014.04.001} {\bibfield  {journal} {\bibinfo
  {journal} {Physics of the Dark Universe}\ }\textbf {\bibinfo {volume}
  {5-6}},\ \bibinfo {pages} {18} (\bibinfo {year} {2014})},\ \Eprint
  {http://arxiv.org/abs/1308.4955} {arXiv:1308.4955 [hep-ph]} \BibitemShut
  {NoStop}%
\bibitem [{\citenamefont {Backovi\'c}\ \emph {et~al.}(2015)\citenamefont
  {Backovi\'c}, \citenamefont {Martini}, \citenamefont {Mattelaer},
  \citenamefont {Kong},\ and\ \citenamefont {Mohlabeng}}]{Backovic:2015cra}%
  \BibitemOpen
  \bibfield  {author} {\bibinfo {author} {\bibfnamefont {M.}~\bibnamefont
  {Backovi\'c}}, \bibinfo {author} {\bibfnamefont {A.}~\bibnamefont {Martini}},
  \bibinfo {author} {\bibfnamefont {O.}~\bibnamefont {Mattelaer}}, \bibinfo
  {author} {\bibfnamefont {K.}~\bibnamefont {Kong}}, \ and\ \bibinfo {author}
  {\bibfnamefont {G.}~\bibnamefont {Mohlabeng}},\ }\href {\doibase
  10.1016/j.dark.2015.09.001} {\bibfield  {journal} {\bibinfo  {journal} {Phys.
  Dark Univ.}\ }\textbf {\bibinfo {volume} {9-10}},\ \bibinfo {pages} {37}
  (\bibinfo {year} {2015})},\ \Eprint {http://arxiv.org/abs/1505.04190}
  {arXiv:1505.04190 [hep-ph]} \BibitemShut {NoStop}%
\bibitem [{\citenamefont {Ambrogi}\ \emph {et~al.}(2019)\citenamefont
  {Ambrogi}, \citenamefont {Arina}, \citenamefont {Backovic}, \citenamefont
  {Heisig}, \citenamefont {Maltoni}, \citenamefont {Mantani}, \citenamefont
  {Mattelaer},\ and\ \citenamefont {Mohlabeng}}]{Ambrogi:2018jqj}%
  \BibitemOpen
  \bibfield  {author} {\bibinfo {author} {\bibfnamefont {F.}~\bibnamefont
  {Ambrogi}}, \bibinfo {author} {\bibfnamefont {C.}~\bibnamefont {Arina}},
  \bibinfo {author} {\bibfnamefont {M.}~\bibnamefont {Backovic}}, \bibinfo
  {author} {\bibfnamefont {J.}~\bibnamefont {Heisig}}, \bibinfo {author}
  {\bibfnamefont {F.}~\bibnamefont {Maltoni}}, \bibinfo {author} {\bibfnamefont
  {L.}~\bibnamefont {Mantani}}, \bibinfo {author} {\bibfnamefont
  {O.}~\bibnamefont {Mattelaer}}, \ and\ \bibinfo {author} {\bibfnamefont
  {G.}~\bibnamefont {Mohlabeng}},\ }\href {\doibase 10.1016/j.dark.2018.11.009}
  {\bibfield  {journal} {\bibinfo  {journal} {Phys. Dark Univ.}\ }\textbf
  {\bibinfo {volume} {24}},\ \bibinfo {pages} {100249} (\bibinfo {year}
  {2019})},\ \Eprint {http://arxiv.org/abs/1804.00044} {arXiv:1804.00044
  [hep-ph]} \BibitemShut {NoStop}%
\bibitem [{\citenamefont {Di~Mauro}\ \emph {et~al.}(2023)\citenamefont
  {Di~Mauro}, \citenamefont {Arina}, \citenamefont {Fornengo}, \citenamefont
  {Heisig},\ and\ \citenamefont {Massaro}}]{DiMauro:2023tho}%
  \BibitemOpen
  \bibfield  {author} {\bibinfo {author} {\bibfnamefont {M.}~\bibnamefont
  {Di~Mauro}}, \bibinfo {author} {\bibfnamefont {C.}~\bibnamefont {Arina}},
  \bibinfo {author} {\bibfnamefont {N.}~\bibnamefont {Fornengo}}, \bibinfo
  {author} {\bibfnamefont {J.}~\bibnamefont {Heisig}}, \ and\ \bibinfo {author}
  {\bibfnamefont {D.}~\bibnamefont {Massaro}},\ }\href {\doibase
  10.1103/PhysRevD.108.095008} {\bibfield  {journal} {\bibinfo  {journal}
  {Phys. Rev. D}\ }\textbf {\bibinfo {volume} {108}},\ \bibinfo {pages}
  {095008} (\bibinfo {year} {2023})},\ \Eprint
  {http://arxiv.org/abs/2305.11937} {arXiv:2305.11937 [hep-ph]} \BibitemShut
  {NoStop}%
\bibitem [{\citenamefont {Amoroso}\ \emph {et~al.}(2019)\citenamefont
  {Amoroso}, \citenamefont {Caron}, \citenamefont {Jueid}, \citenamefont
  {Ruiz~de Austri},\ and\ \citenamefont {Skands}}]{Amoroso:2018qga}%
  \BibitemOpen
  \bibfield  {author} {\bibinfo {author} {\bibfnamefont {S.}~\bibnamefont
  {Amoroso}}, \bibinfo {author} {\bibfnamefont {S.}~\bibnamefont {Caron}},
  \bibinfo {author} {\bibfnamefont {A.}~\bibnamefont {Jueid}}, \bibinfo
  {author} {\bibfnamefont {R.}~\bibnamefont {Ruiz~de Austri}}, \ and\ \bibinfo
  {author} {\bibfnamefont {P.}~\bibnamefont {Skands}},\ }\href {\doibase
  10.1088/1475-7516/2019/05/007} {\bibfield  {journal} {\bibinfo  {journal}
  {JCAP}\ }\textbf {\bibinfo {volume} {05}},\ \bibinfo {pages} {007} (\bibinfo
  {year} {2019})},\ \Eprint {http://arxiv.org/abs/1812.07424} {arXiv:1812.07424
  [hep-ph]} \BibitemShut {NoStop}%
\end{thebibliography}%

\clearpage
\maketitle
\onecolumngrid
\begin{center}
%\textbf{\large CosmiXs: spectra of cosmic antideuterons from dark matter interactions} \\ 
%\vspace{0.05in}
{\bf \large SUPPLEMENTAL MATERIAL}
%{Mattia Di Mauro, Nicolao Fornengo, Adil Jueid, Roberto Ruiz de Austri, Chiara Arina, and Francesca Bellini}
\vspace{0.05in}
\end{center}
\onecolumngrid
%%%%%%%%%% Merge with Supplemental material %%%%%%%%%%
\setcounter{equation}{0}
\setcounter{figure}{0}
\setcounter{table}{0}
\setcounter{section}{0}
\setcounter{page}{1}
\makeatletter
\renewcommand{\theequation}{S\arabic{equation}}
\renewcommand{\thefigure}{S\arabic{figure}}

\section{Wigner formalism}
\label{app:wigner}

The Wigner formalism allows for a full quantum-mechanical treatment of the coalescence process, which takes into account the momentum distribution of the nucleons, the nucleus wave function, and the characteristics of the nucleon emitting source~\cite{Bellini:2018epz, Blum:2017qnn, Bellini:2020cbj, Kachelriess:2019taq}. 
In this formalism, the differential $\overline{\rm D}$ spectrum is obtained through a projection of the two-nucleon density matrix $\rho_{\mathrm{pn}}$ on the $\overline{\rm D}$ density matrix $\rho_{\mathrm{d}}$:
$$
\frac{\mathrm{d}^3 N_{\mathrm{d}}}{\mathrm{d} p_{\mathrm{d}}^3}=\operatorname{tr}\left(\rho_{\mathrm{d}} \rho_{\mathrm{pn}}\right).
$$
The density matrices are defined as $\rho_{\mathrm{d}}=\left|\phi_{\mathrm{d}}\right\rangle\left\langle\phi_{\mathrm{d}}\right|$ and $\rho_{\mathrm{pn}}=\left|\psi_{\mathrm{pn}}\right\rangle\left\langle\psi_{\mathrm{pn}}\right|$, 
where $\phi_{\mathrm{d}}$ and $\psi_{\mathrm{pn}}$ are the antideuteron wavefunction and the two-particle $\bar{p}$--$\bar{n}$ wavefunction, respectively. We can factorize the spatial ($\bf{r}$) and momentum ($\bf{p}$) dependencies of the wave function $\phi_{\mathrm{d}}$ and write it as:
$$
\phi_{\mathrm{d}}\left({\bf r}, {\bf p}\right) \propto \varphi_{\mathrm{d}} ({\bf r}) e^{i {\bf p} \cdot {\bf r}_{\mathrm{d}}},
$$
where $\varphi_{\mathrm{d}}$ is the internal $\overline{\rm D}$ wavefunction, $\bf{r}_{\mathrm{d}}$ is the space-time position of the deuteron and $\bf{p}$ its four-momentum. Taking this into account, the antideuteron spectrum takes the form \cite{Scheibl:1998tk,Kachelriess:2019taq}:
\begin{equation}
\label{eq:wignerDspectrum}
\frac{\mathrm{d}^3 N_{\mathrm{d}}}{\mathrm{d} p_{\mathrm{d}}^3}= S \int \frac{\mathrm{d}^3 r_{\mathrm{d}} \mathrm{d}^3 r \mathrm{~d}^3 q}{(2 \pi)^6} \cdot  \mathcal{D}(\vec{r}, \vec{q})
\cdot W_{\mathrm{pn}}\left(\vec{p}_{\mathrm{d}} / 2+\vec{q}, \vec{p}_{\mathrm{d}} / 2-\vec{q}, \vec{r}_{\mathrm{p}}, \vec{r}_{\mathrm{n}}\right),
\end{equation}
where $S$ is a factor that takes into account spin and isospin statistics, which is equal to $3 / 8$ for a $\overline{\rm D}$. $\vec{r}_{\mathrm{p}}$ and $\vec{r}_{\mathrm{n}}$ are the proton and neutron positions, $\vec{r} \equiv \vec{r}_{\mathrm{p}}-\vec{r}_{\mathrm{n}}$, and $\vec{q}$ is $\vec{q}\equiv\left(\vec{p}_{\mathrm{p}}-\vec{p}_{\mathrm{n}}\right) / 2 = \Delta \vec{p}/2$, where $\vec{p}_{\mathrm{p}}$ and $\vec{p}_{\mathrm{n}}$ are the antiproton and antineutron momentum, respectively. 

The Wigner function of the $\bar{p}$--$\bar{n}$ state is $W_{\mathrm{pn}}$ and $\mathcal{D}$ is the Wigner function of the antideuteron, defined as \cite{Scheibl:1998tk}:
$$
\mathcal{D}(\vec{r}, \vec{q})=\int \mathrm{d}^3 \xi e^{-i \vec{q} \cdot \vec{\xi}} \varphi_{\mathrm{d}}(\vec{r}+\vec{\xi} / 2) \varphi_{\mathrm{d}}^*(\vec{r}-\vec{\xi} / 2) .
$$
The antideuteron Wigner function is normalized in such a way that:
\begin{equation}
\label{eq:Wignormalized}
\int d^3r \int \frac{d^3q}{(2 \pi)^3}\mathcal{D}(\vec{r}, \vec{q}) = 1 .
\end{equation}

The choice of the shape for the $\overline{\rm D}$ wavefunction $\varphi_{\mathrm{d}}$ affects the form of the antideuteron Wigner function $\mathcal{D}(\vec{r}, \vec{q})$ and we will adopt two different options, discussed in the next Sections.

Concerning the $\bar{p}$--$\bar{n}$ Wigner function $W_{\mathrm{pn}}$, we   factorize the space and momentum dependences as:
$$
W_{\mathrm{pn}}=H_{\mathrm{pn}}\left(\vec{r}_{\mathrm{p}}, \vec{r}_{\mathrm{n}}\right) G_{\mathrm{pn}}\left(\vec{p}_{\mathrm{d}} / 2+\vec{q}, \vec{p}_{\mathrm{d}} / 2-\vec{q}\right),
$$
where $G_{\mathrm{pn}}$ is the two-particle momentum distribution, that we obtain from a \textsc{Pythia}~8 modeling of the $\bar{p}$ and $\bar{n}$ production processes, which includes the nucleon single-particle momentum distributions and their correlations. 
For the space term $H_{\mathrm{pn}}$, the $\bar{n}$ and $\bar{p}$ momentum distribution can be separated:
$$
H_{\mathrm{pn}}\left(\vec{r}_{\mathrm{p}}, \vec{r}_{\mathrm{n}}\right)=h\left(\vec{r}_{\mathrm{p}}\right) h\left(\vec{r}_{\mathrm{n}}\right),
$$
where $h\left(\vec{r}_{\mathrm{p}}\right)$ and $h\left(\vec{r}_{\mathrm{n}}\right)$ are the spatial single-particle distributions and are also obtained from the same \textsc{Pythia}~8 modeling.

The specific implementation of the Wigner formalism for the calculation of the antideuteron spectrum therefore depends on the assumption done for the antinucleons and antideuteron wave functions and the momenta and spatial distributions of the coalescing antinucleons.

\subsection{Gaussian wavefunction}
\label{sec:argonne}

Our first model assumes a Gaussian form for the antideuteron internal wavefunction:
\begin{equation}
    \varphi_{\rm{d}}(r) = \left(\pi d^2\right)^{-3/4} e^{-r^2/(2d^2)},
    \label{eq:deuteron_wf_Gaus}
\end{equation}
where $r$ represents relative distance between the antinuclei such that the deuteron RMS charge radius is calculated as $r_{RMS} = \int dr^3 (r/2)^2|\varphi_d(r)|^2$. 
Instead, $d$ is the deuteron size parameter of the wavefunction. In particular, a value of $d=3.2$ fm reproduces the measured $\overline{\rm D}$ RMS charge radius \cite{CREMA:2016idx}.
We assume a Gaussian shape also for the $\bar{p}$ and $\bar{n}$ spatial distribution functions $h\left(\vec{r}_{\mathrm{p}}\right)$ and $h\left(\vec{r}_{\mathrm{n}}\right)$:
\begin{equation}
    h(r_{p/n}) = \left( 2 \pi \sigma^2\right)^{-3/2} e^{- r_{p/n}^2/(2\sigma^2)},
    \label{eq:deuteron_wf_Gaus}
\end{equation}
where $\sigma$ is the size of the two-particle emitting source.
As a result, the corresponding antideuteron Wigner function takes the form \cite{Kachelriess:2020uoh}
\begin{equation}
     \mathcal{D}(r,q) = 8e^{-r^2/d^2}e^{-q^2 d^2}.
     \label{eq:deutwignerinternal}
\end{equation}
Previous studies, such as \cite{Kachelriess:2019taq,Horst:2023oti}, have integrated out the spatial information of the antinucleon and antideuteron wave functions by performing the integral in Eq.~(\ref{eq:wignerDspectrum}) with respect to $\vec{r}$ and $\vec{r}_d$. By following this approach and assuming Gaussian shapes for the wave functions, the $\overline{\rm D}$ spectrum is given by:
\begin{equation}
    \frac{d^{3}{N_d}}{dp_d^3} = \frac{3}{(2\pi)^6} \left(\frac{d^2}{d^2 + 4\sigma^2}\right)^{3/2} 
    \int d^{3}{q}\,e^{-q^2 d^2}\,
     G_{np}(\vec{p}_{\rm{d}},\vec{q}).
    \label{eq:singlegaussian}
\end{equation}
The pre-factor in front of the integral depends only on the deuteron size $d$ and on the spatial spread $\sigma$ of the nucleon source.
The antideuteron spectrum is thus calculated by taking the $\bar{n}$ and $\bar{p}$ momenta distribution $G_{np}$ from the MC event generator and by calculating Eq.~(\ref{eq:singlegaussian}) numerically.

However, the result presented in Eq.~(\ref{eq:singlegaussian}), does not specifically take into account the spatial and momentum correlations between antinucleons.
Instead, in our paper we have evaluated also the spatial distribution of $\bar{p}$ and $\bar{n}$ directly from \textsc{Pythia}. 
In order to do so, we have assumed the Wigner function to be a PDF, which depends on the distance $\Delta \vec{r}$ and difference of momenta $\Delta \vec{p}$ of antinuclei.
When using Gaussian shapes for the Wigner function the PDF is defined as: 
\begin{equation}
    {\rm PDF}(r,q) = N e^{-r^2/(2\sigma^2)}e^{-q^2 \delta^2/2},
    \label{eq:singlegaussianPDF}
\end{equation}
where $N$ is fixed in such a way that the PDF is correctly normalized and the widths of the Gaussian function in space and momentum are parametrized with the $\sigma$ and $\delta$ parameter.
In the Wigner formalism $\sigma$ and $\delta$ values should be the same and of the order of $d/\sqrt{2} \approx 2$ fm (see Eq.~\ref{eq:deutwignerinternal} and \ref{eq:singlegaussianPDF}).
We can define the coalescence probability $\mathcal{P}$ to form a $\overline{\rm D}$ from two antinucleons with a given distance $\vec{r}$ and momentum difference $\vec{q}$ as follows
\begin{equation}
\label{eq:coalProb}
\mathcal{P}(r, q)= \int_0^r \int_0^q  dr \, dq \, \mathcal{D}(r, q) .
\end{equation}
This probability is correctly normalized as written in Eq.~(\ref{eq:Wignormalized}).

\subsection{Argonne wavefunction}
\label{sec:gaussian}

The Argonne $v_{18}$ potential is a phenomenological potential for the deuteron, tuned through $p–p$ and $n–p$ inelastic scattering data, low-energy $n-n$ scattering parameters, and deuteron binding energy \cite{Wiringa:1994wb}. In such a potential, the deuteron wavefunction has the form:
\begin{equation}
\varphi_{\mathrm{d}}(\vec{r})=\frac{1}{\sqrt{4 \pi} r}\left[u(r)+\frac{1}{\sqrt{8}} w(r) S_{12}(\hat{r})\right] \chi_{1 m},
\end{equation}
where $S_{12}(\hat{r})=3 \,(\vec{\sigma}_1 \cdot \vec{r}) \, (\vec{\sigma}_2 \cdot \vec{r})-(\vec{\sigma}_1 \cdot \vec{\sigma}_2)$ is the spin tensor, $\sigma_i$ are the Pauli matrices, $\chi_{1 m}$ is a spinor, and $u(r)$ and $w(r)$ are radial $\mathrm{S}$ and $\mathrm{D}$ wavefunctions, respectively.
The wavefunction is normalised as follows:
$$
\int \mathrm{d}^3 r\left|\varphi_{\mathrm{d}}(\vec{r})\right|^2=\int \mathrm{d}^3 r \frac{1}{4 \pi r^2}\left[u^2(r)+w^2(r)\right]=1 .
$$
We have tabulated the Wigner function $\mathcal{D}(r,q)$ found with the Argonne wavefunction as a function of $r$ and $q$ following the fit presented in \cite{Horst:2023oti}.

\section{Spectra of $\overline{\rm D}$}
\label{sec:appspectra}

\begin{figure*}
\includegraphics[width=0.329\linewidth]{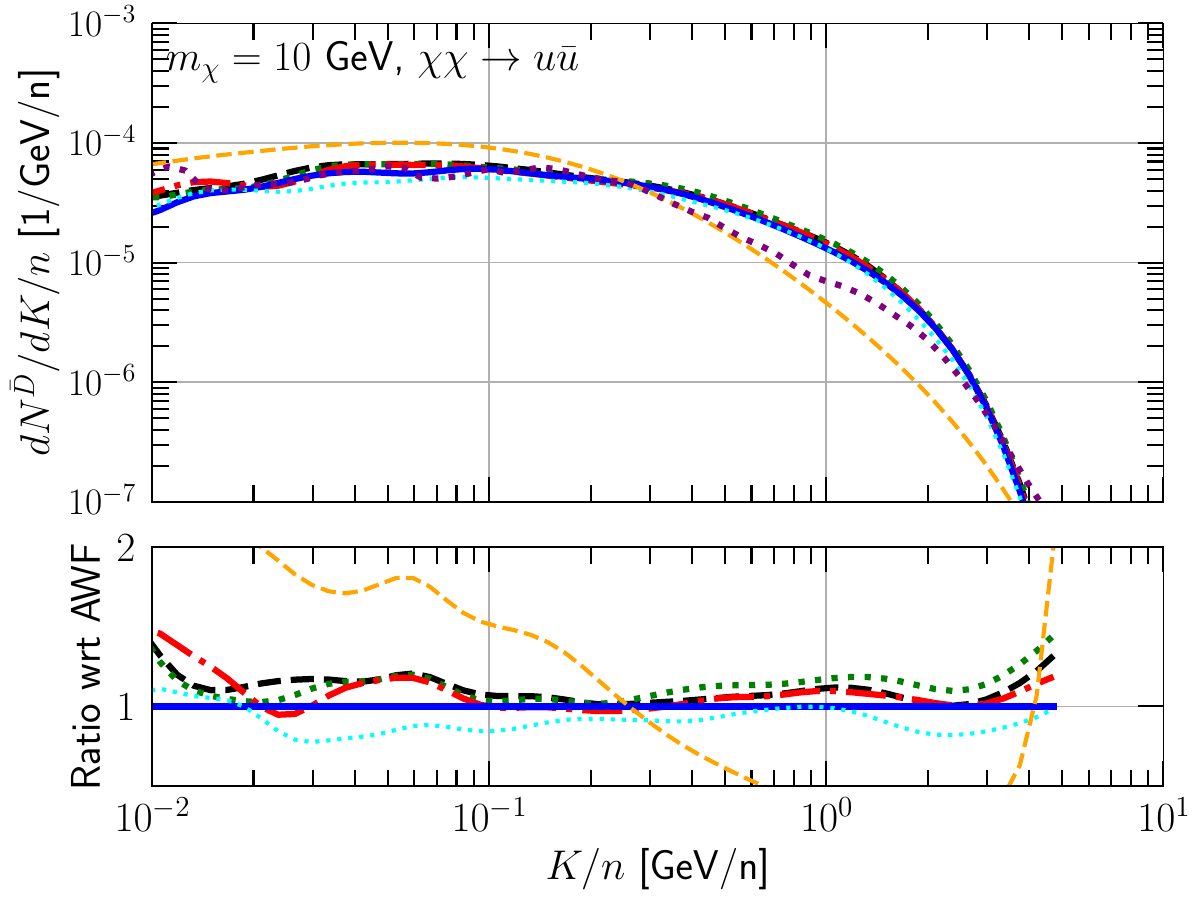}
\includegraphics[width=0.329\linewidth]{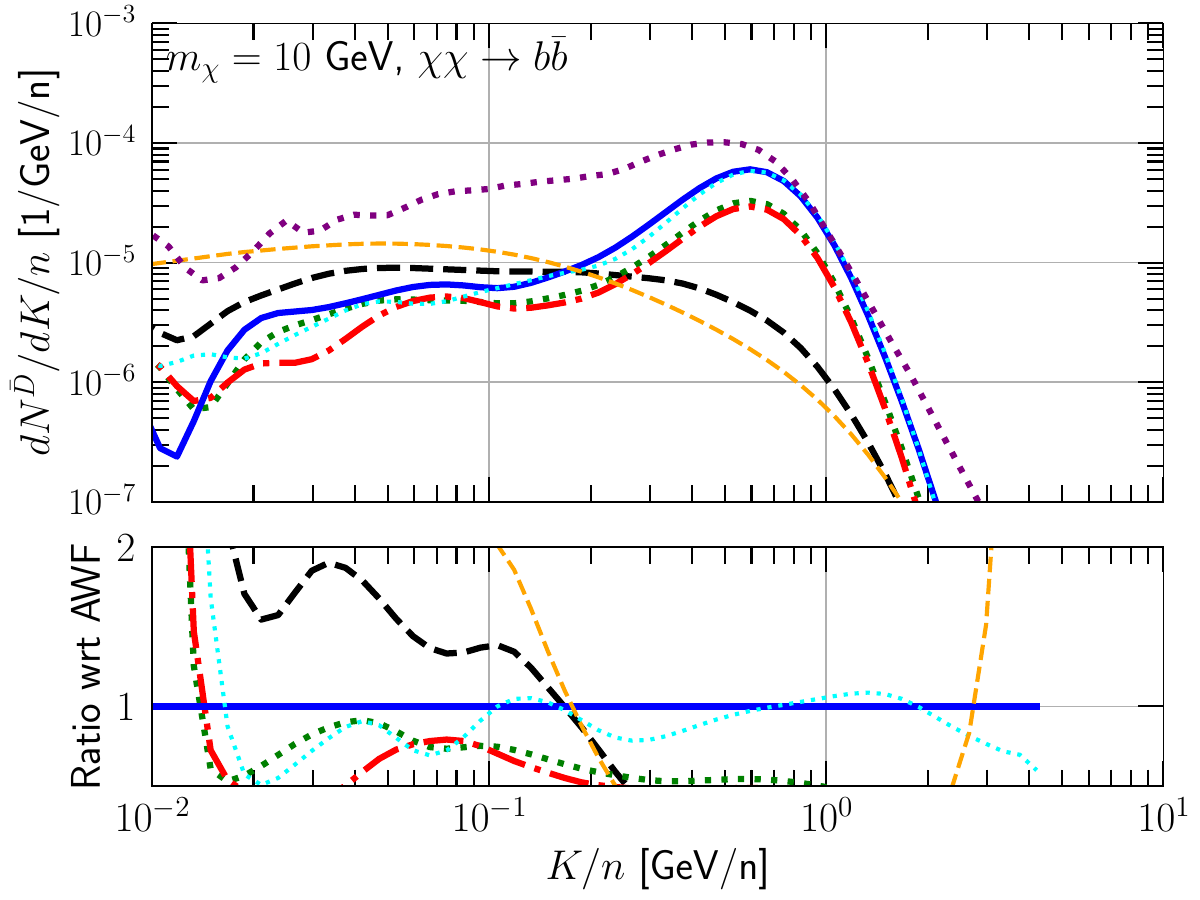}
\includegraphics[width=0.329\linewidth]{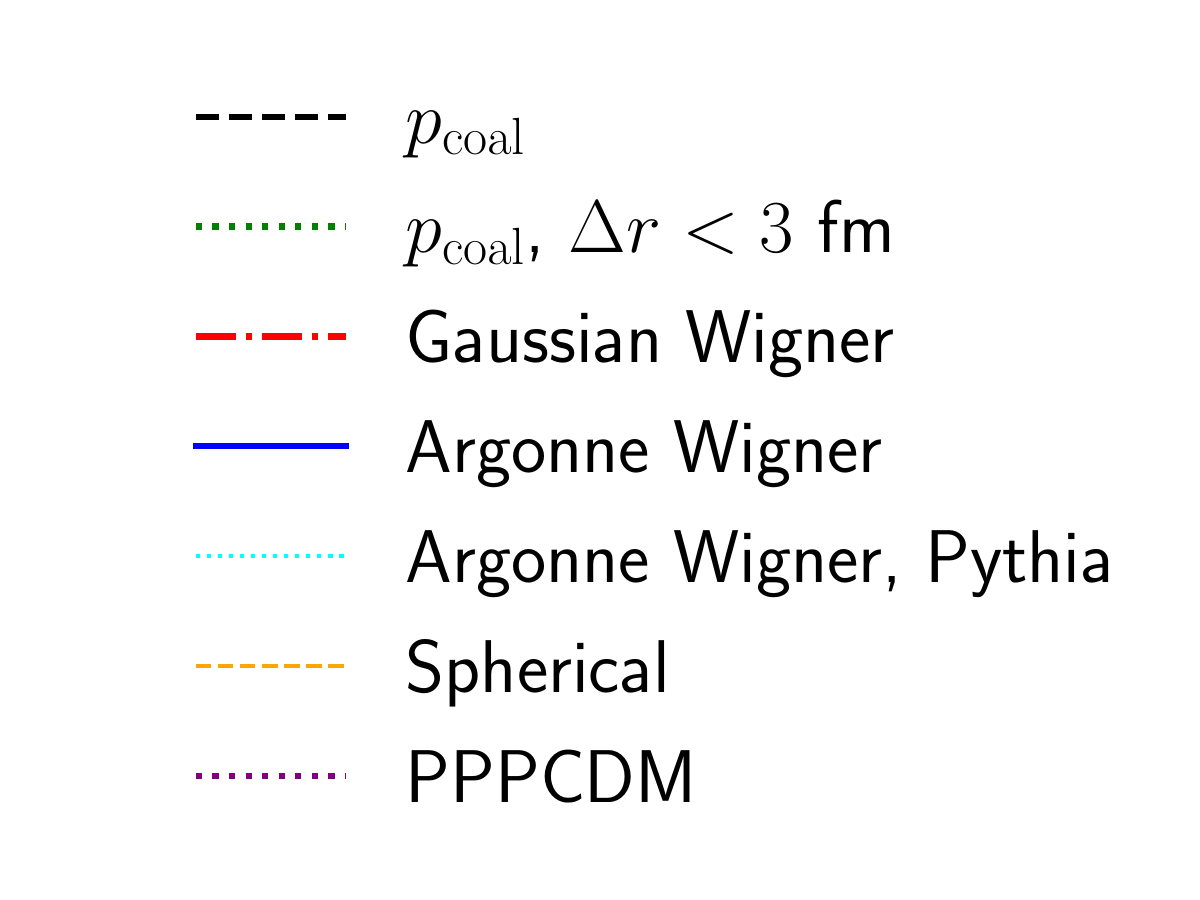}
\includegraphics[width=0.329\linewidth]{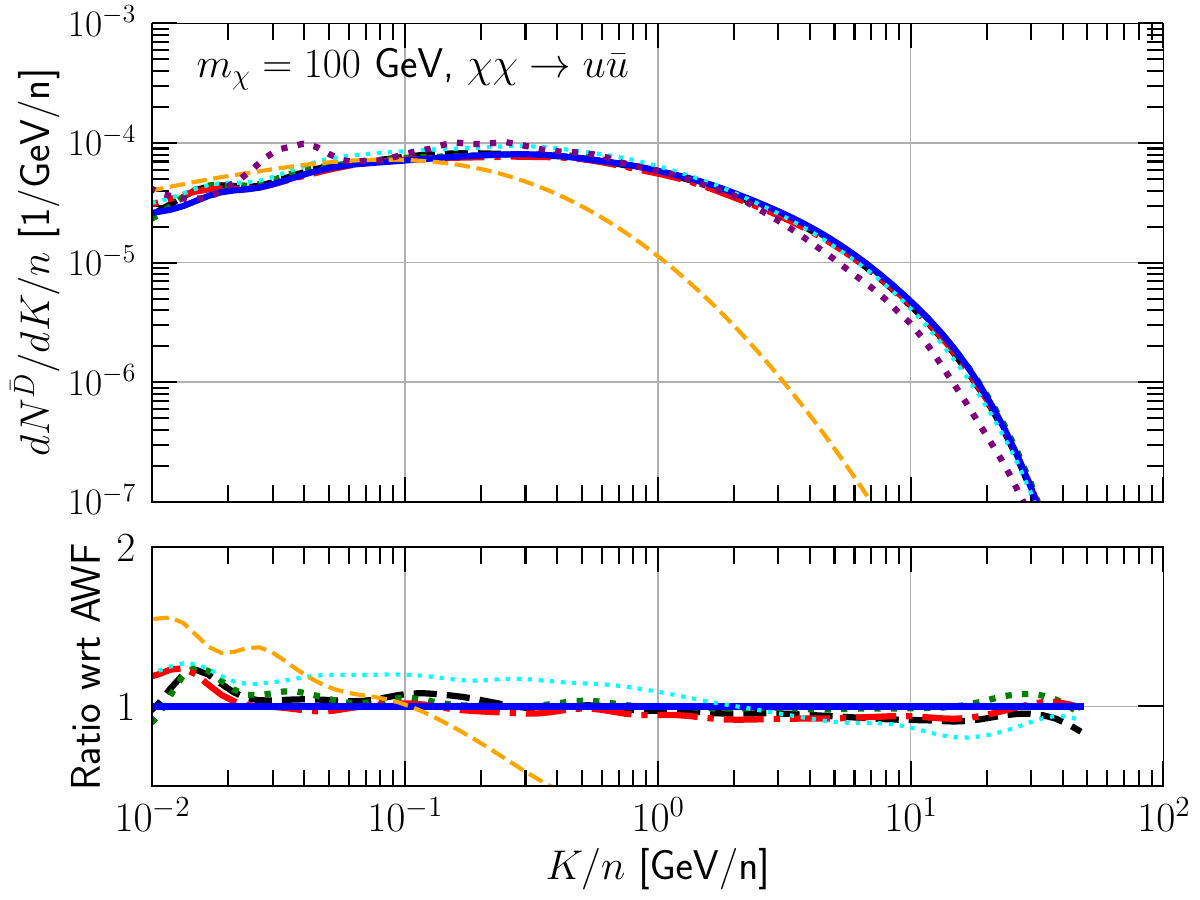}
\includegraphics[width=0.329\linewidth]{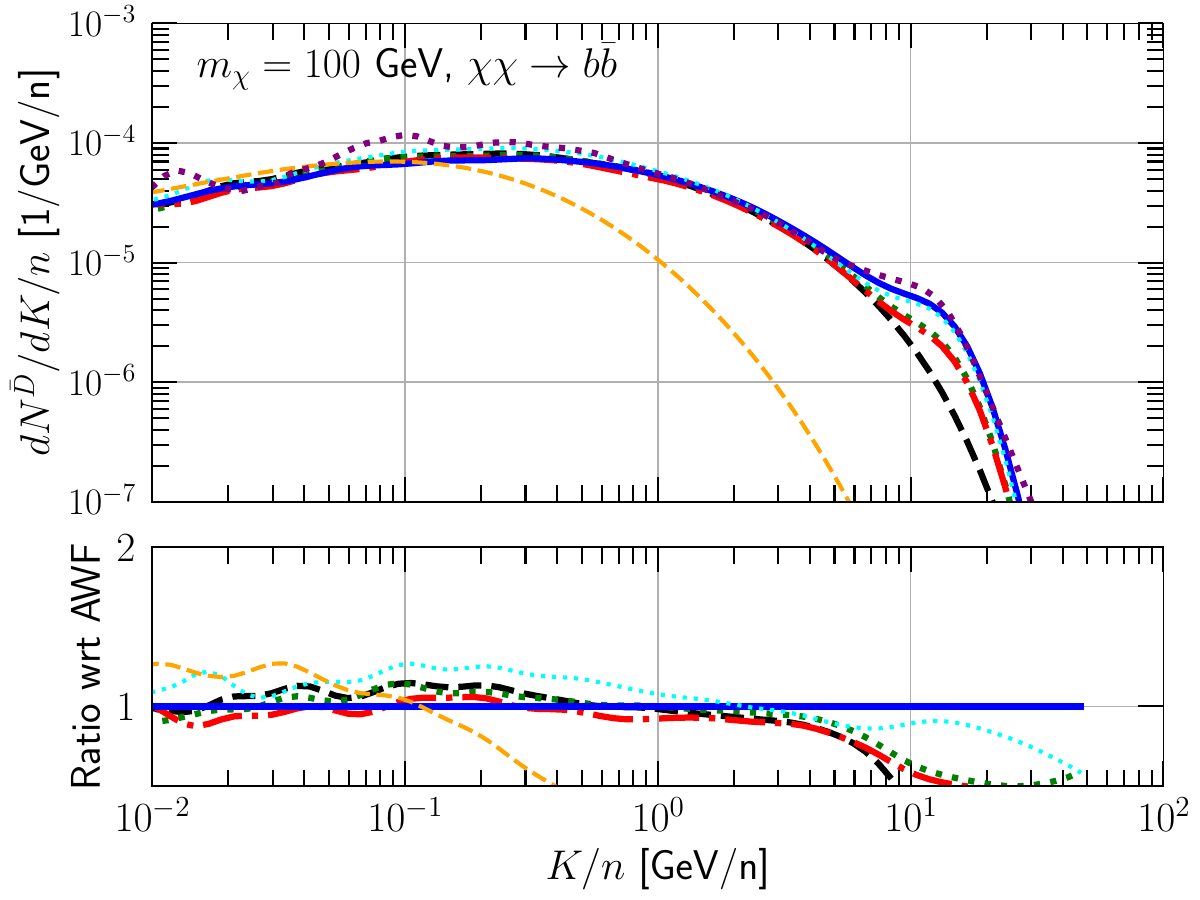}
\includegraphics[width=0.329\linewidth]{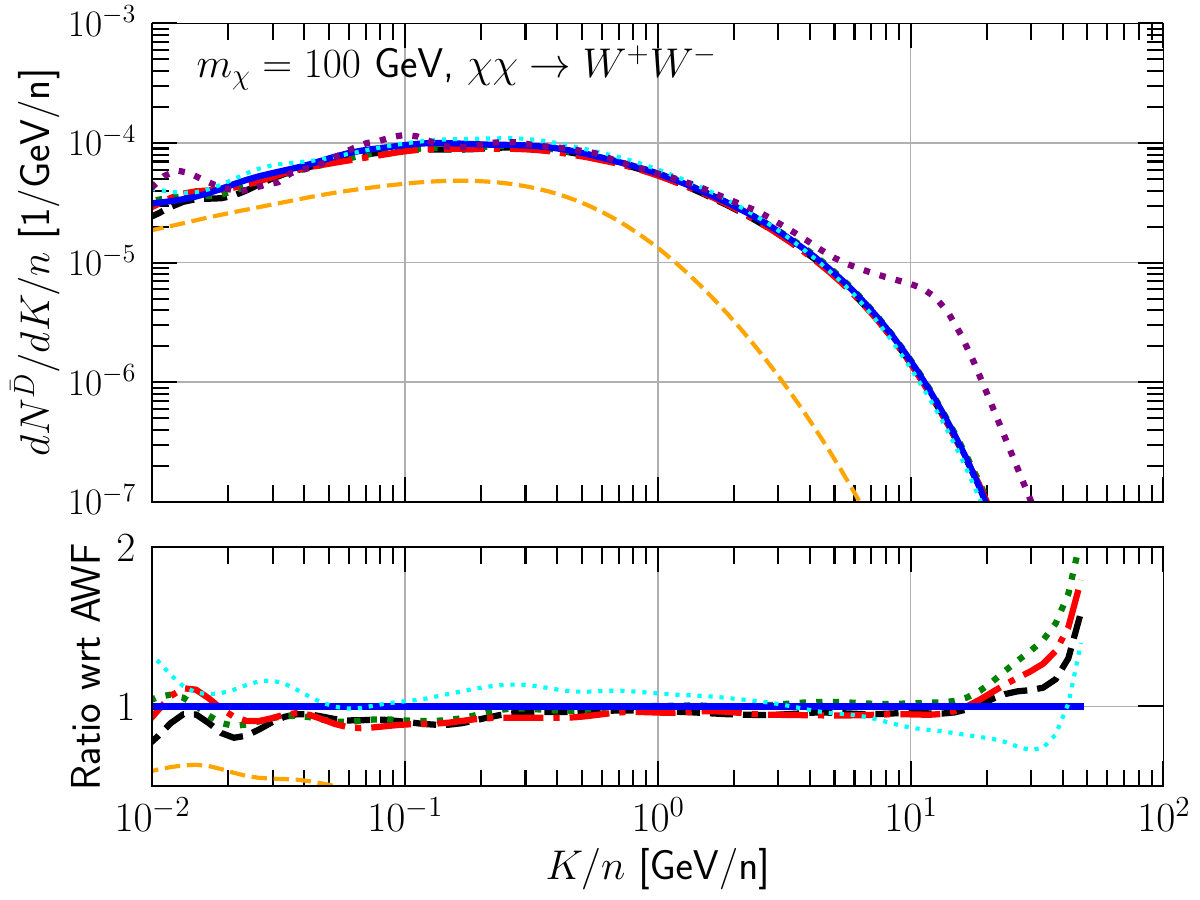}
\includegraphics[width=0.329\linewidth]{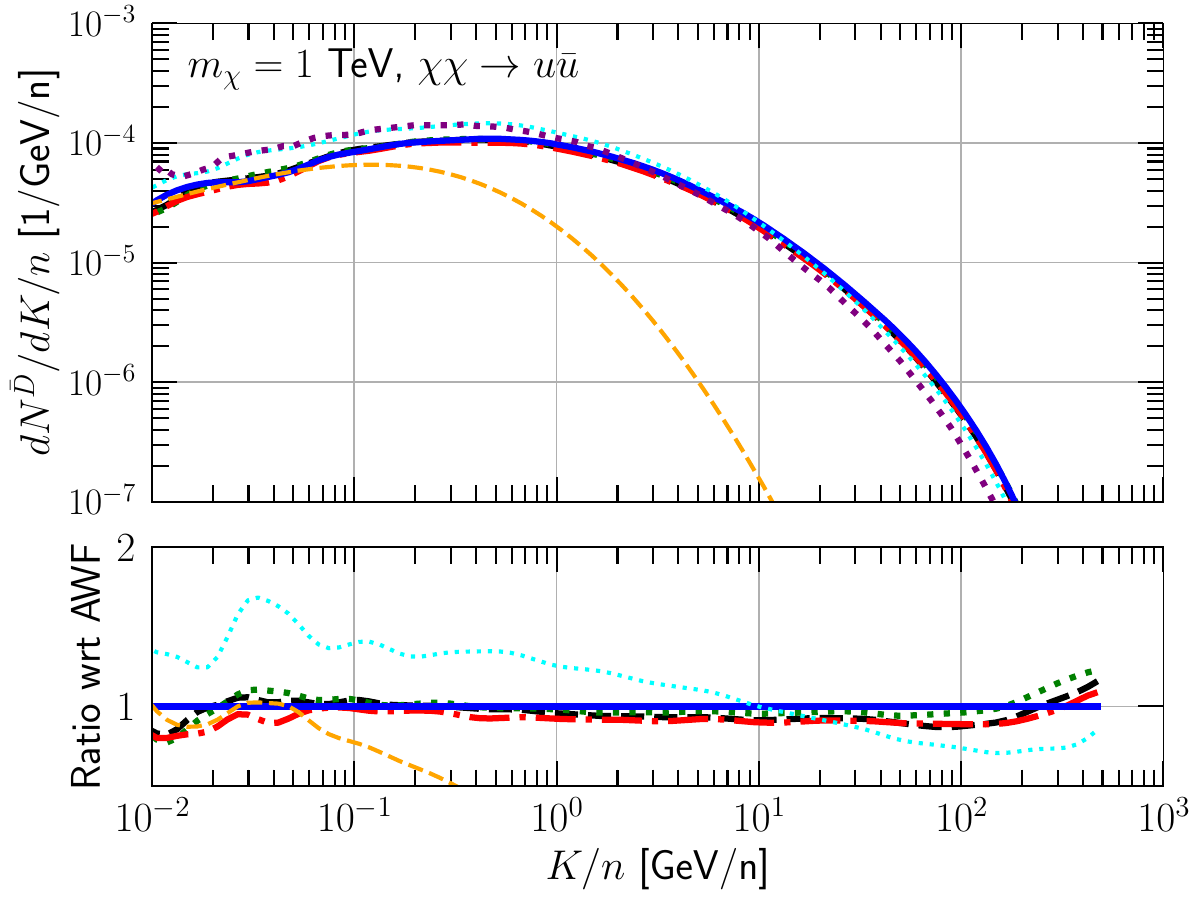}
\includegraphics[width=0.329\linewidth]{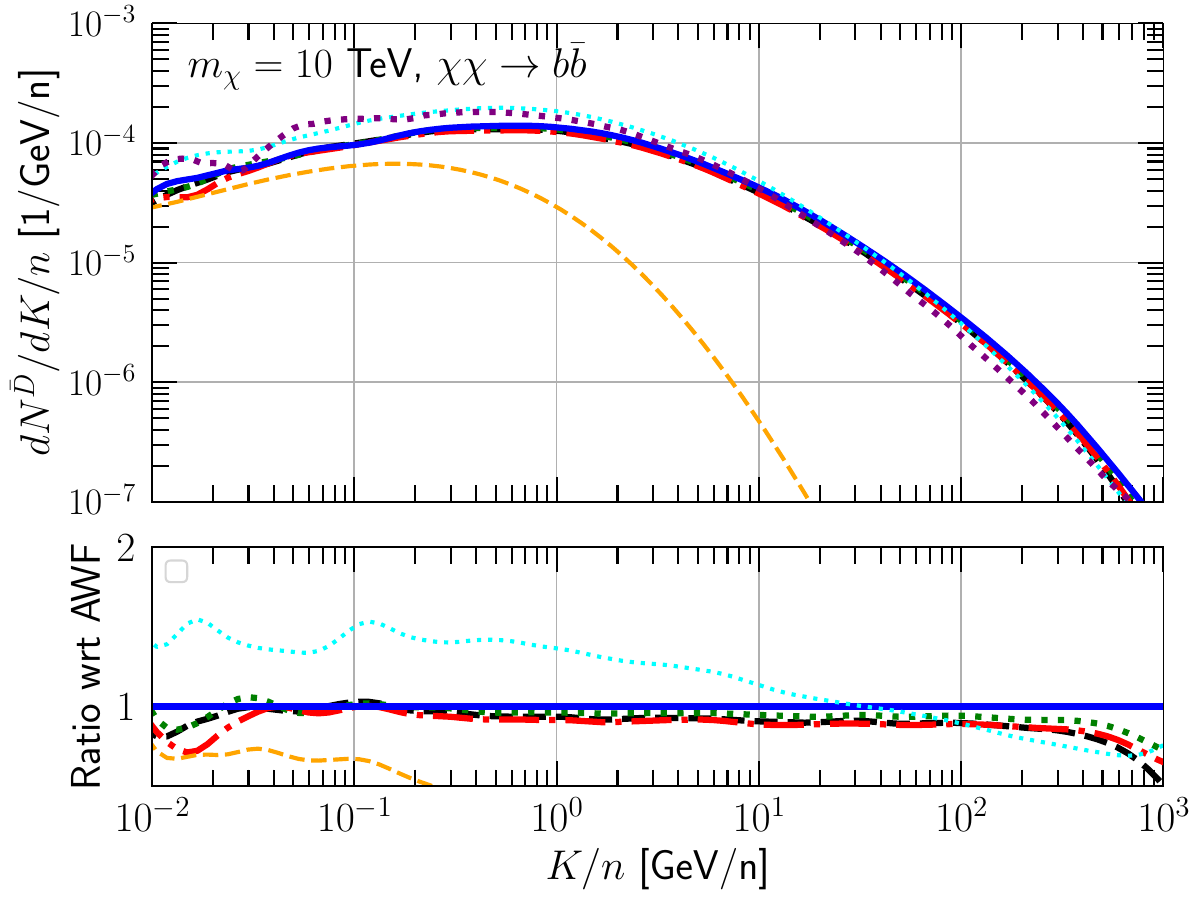}
\includegraphics[width=0.329\linewidth]{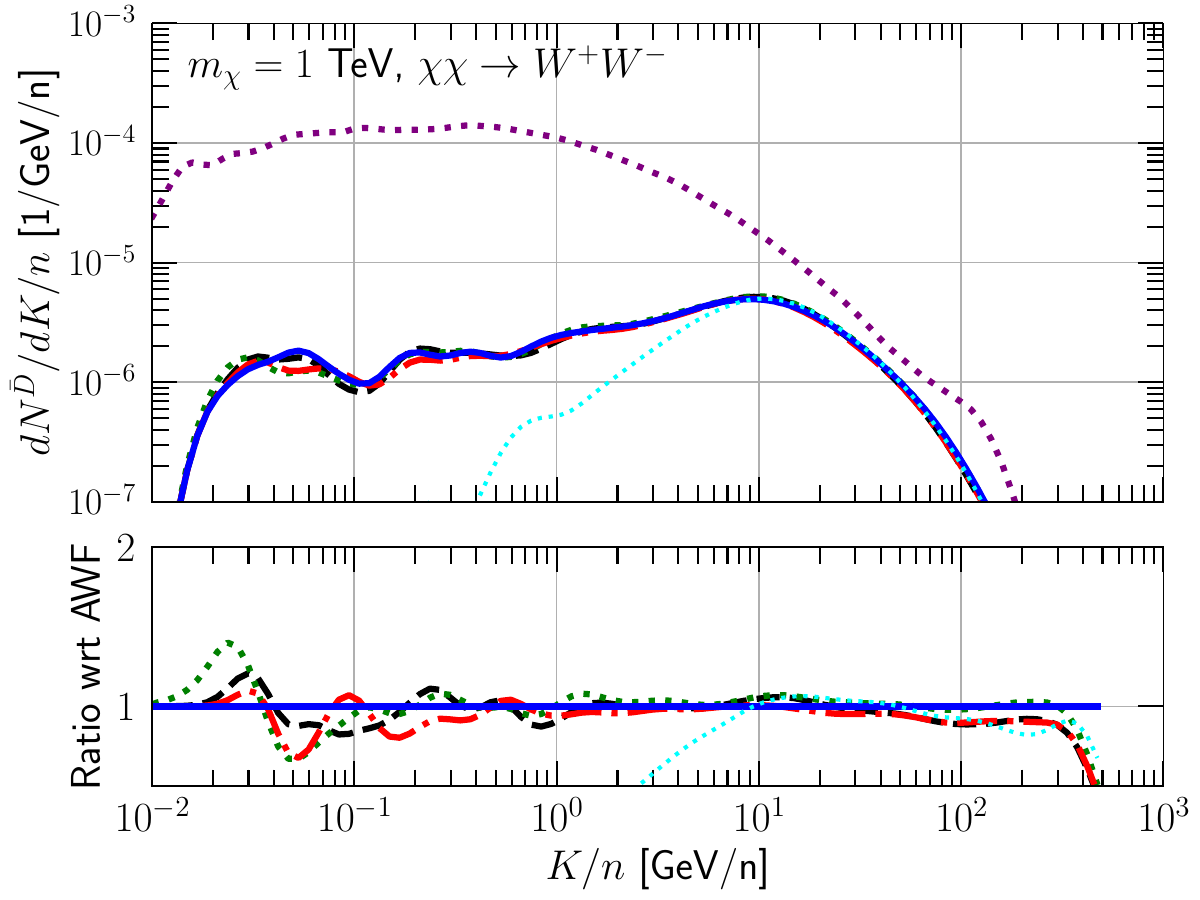}
\includegraphics[width=0.329\linewidth]{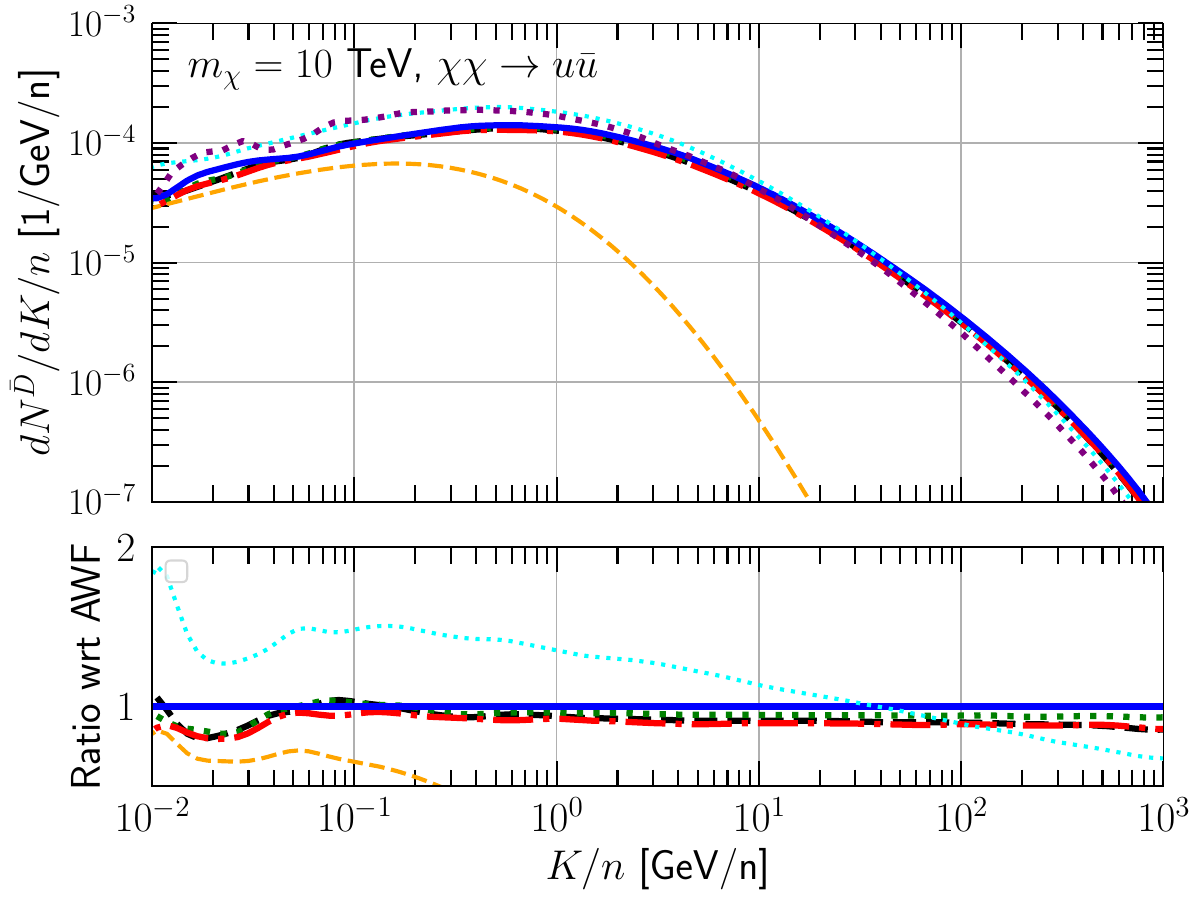}
\includegraphics[width=0.329\linewidth]{figures/dNdKn_ratio_bb_10000GeV_allmethods_paper.pdf}
\includegraphics[width=0.329\linewidth]{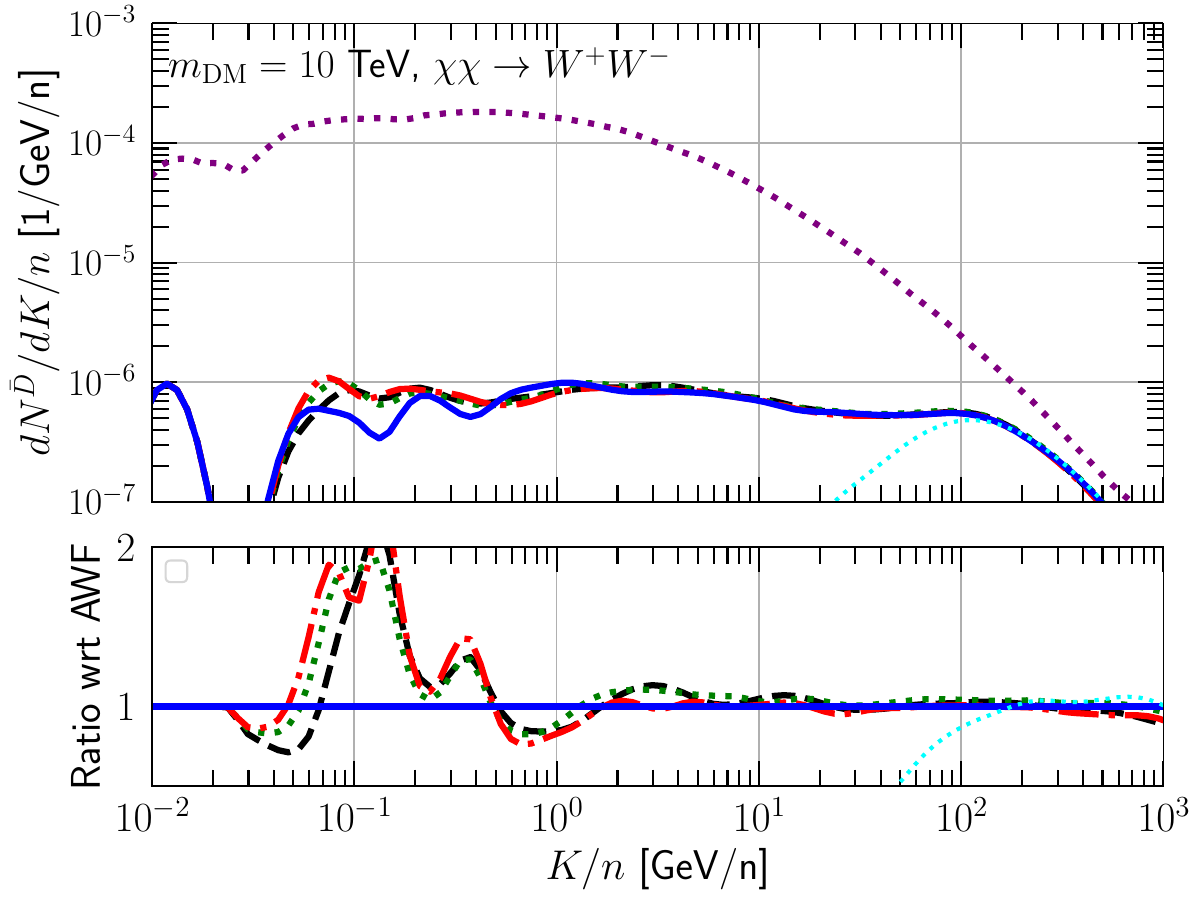}
    \caption{Spectra of $\overline{\rm D}$ as a function of kinetic energy per nucleon, for the $u\bar{u}$ (left panel), $b\bar{b}$ (central panel) and $W^+W^-$ (right panel) annihilation channels. We show the results for different masses, from top to bottom from 10 GeV up to $m_{\rm{DM}}= 10$ TeV. The spectra are obtained using \textsc{Pythia}~8.309 with the \textsc{Vincia} shower algorithm, for all the coalescence models discussed in the paper. The case denoted Argonne Wigner Pythia refers to the adoption of the standard \textsc{Pythia} shower algorithm instead of \textsc{Vincia}.
    For comparison, we also display the spectra calculated with PPPCDM. The lower panels show the ratio between the spectra obtained in different models with the Argonne Wigner model. \cite{Cirelli:2010xx}.}
    \label{fig:spectrall}
\end{figure*}

In this section we report further results for production of antideuterons from DM annihilation.
In particular, in Fig.~\ref{fig:spectrall} we show the source spectra evaluated as a function of the kinetic energy over number of nucleons for three different annihilation channels: $u\bar{u}$, $b\bar{b}$ and $W^+W^-$. The spectra obtained for the other quark and boson channels are similar to the cases shown here.
Instead, the leptonic channels provide an antideuteron yield that is several orders of magnitude smaller. For example the $\tau^+\tau^-$ channel, which is the one with a larger production of hadrons, gives a multiplicity that is a factor of about 100 smaller than the $u\bar{u}$ channel.  

As already noted in the main text, the difference of the spectra, obtained when using different coalescence models, can be large if the DM masses in below a few tens of GeV. In fact, in the case with $m_{\rm{DM}}=10$ GeV the spectra found with the $u\bar{u}$ channel vary by $20\%$ while for the $b\bar{b}$ channel even by a factor of 2.
Instead, when increasing the DM mass the simple coalescence and Wigner formalism models provide spectra that are consistent within about $5-10\%$ in the relevant energies for DM detection.

The spherical approach can give similar results at very low kinetic energies while above $1 GeV/n$ it typically underestimate the antideuteron yield. This has been already noted, e.g.~ in Ref.~\cite{Fornengo:2013osa}, and it is due to the presence of correlations in the antinucleons phase space which are taken into account in the MC appraoch and are not considered in the spherical model. These correlations become progressively more relevant when the two quarks are produced at larger energies. In this case, the two back-to-back emerging jets are more focused and the ensuing antinucleons possess, on average, closer relative momenta and therefore more antideuterons large energies are produced.

By looking to Fig.~\ref{fig:spectrall} we can also appreciate the difference obtained when using the standard Pythia or the new VINCIA shower algorithm. The differences become particularly relevant increasing the DM mass. In fact, VINCIA includes a much more refined treatment of the EW corrections that become important at high energies.
For the quark channels and DM masses above 1 TeV the effect of the improved EW treatment in VINCIA leads to differences with respect to Pythia results of the order of $20\%-30\%$. Instead, for the gauge boson channels the effect becomes much more dramatic. In fact, for these channels an incomplete or wrong treatment of the EW corrections can lead to large differences for kinetic energies below 10 GeV/n.

We also note that the results we obtain are similar to the ones reported in Ref.~\cite{Cirelli:2010xx} for the quark channels. Instead, for the gauge boson ones and DM masses above 1 TeV the spectra do not match and we deviations that reach even 1-2 orders of magnitude. This is also visible in the Fig.~\ref{fig:mult} where we show the multiplicity. While the multiplicity obtained for all the coalescence models and with Pythia or VINCIA in our case is basically flat as a function of DM mass, the one obtained in Ref.~\cite{Cirelli:2010xx} has an increasing trend. Our findings are consistent with the one obtained in Ref.~\cite{Kadastik:2009ts,Fornengo:2013osa}.

In our analysis, when using coalescence models that include also a criteria based on the distance between antinucleons, 
the antideuterons produced from weakly decaying baryons and mesons are included.
Ref.~\cite{Winkler:2020ltd} claimed that the antinuclei, in particular the anti-helium3, yield can be significantly enhanced due to the decay of the $\bar{\Lambda}_b$. This production channel mostly contributes in the case of $b\bar{b}$ annihilation channel for DM masses below 100 GeV and at energies close to the kinematic cutoff of the DM mass. In Fig.~\ref{fig:spectrall} we can appreciate the $\bar{\Lambda}_b$ contribution as the bump present in the highest-energy part of the spectra for the cases with $m_{\rm{DM}}=10$ and 100 GeV.
However, we do not find a contribution as large as the one reported in Ref.~\cite{Winkler:2020ltd}. This will be extensively discussed in a forthcoming paper which is focused on $\overline{\rm D}$ and $\overline{\rm He}$ production from weakly decaying hadrons.

\section{Monte Carlo setup}
\label{sec:montecarlo}

The technical setup of event generation with Pythia, that is used in this work, follows the one implemented in Ref.~\cite{Arina:2023eic}. We summarize here the main ingredients and refer the reader to Ref.~\cite{Arina:2023eic} for a comprehensive treatment. 

We use the \textsc{Vincia} shower algorithm, which is implemented in the \textsc{Pythia}~8 event generator version 8.309, that we interface to \textsc{MadDM}~\cite{Backovic:2013dpa,Backovic:2015cra,Ambrogi:2018jqj}. 
\textsc{Vincia} represents the state-of-the-art shower algorithm that includes a complete treatment of the EW corrections. In particular, it includes contributions from triple gauge boson interactions which are very important for DM masses above 1 TeV. 
The interface between \textsc{Vincia} and \textsc{MadDM} permits to take into account the helicity information of the particles in the hard process during the hadronic and electromagnetic showers. 
Considering properly the particle helicity allows to calculate correctly the EW corrections and to include off-shell contributions of the gauge bosons.
To this end, we employ as a particle physics model of DM the  Singlet Scalar model with a Higgs portal (see, e.g.~\cite{DiMauro:2023tho}). 

We perform a dedicated tuning of the hadronization parameters in \textsc{Pythia} by fitting the available measurements reported on by ALEPH, DELPHI, L3 and OPAL experiments for the production of mesons and baryons at the $Z$-boson pole (see also Refs. \cite{Amoroso:2018qga,Jueid:2022qjg,Jueid:2023vrb} for more details). 
These parameters are used to form predictions for DM annihilation source spectra thanks to the universality of the hadronization process. For each DM mass, we simulate a number of events that depends on the value of the DM mass. In particular, we chose the number of events is such a way that the statistical uncertainty at the peak of the source spectra is $5\%$. For example, in case of the $b\bar{b}$ annihilation channel and DM mass of 100 GeV, we simulate 200 million events. For lower (higher) masses the number is increased (decreased).

For each \textsc{Pythia} simulation we select all the pairs of $\bar{n}$ and $\bar{p}$ present in the event list and apply the coalescence criteria discussed in the manuscript main text. We then determine the difference of momenta $\Delta p$ and of distance $\Delta r$ calculated in the reference frame of the $\bar{n}$ and $\bar{p}$.
If the coalescence criterium is satisfied for a pair of $\bar{n}$ and $\bar{p}$, we assume that the $\overline{\rm D}$ is formed and we calculate its kinetic energy in the CM of the DM annihilation process. Once, the event simulations is finished, we calculate the spectrum as:
\begin{equation}
\frac{dN}{dK}_i = \frac{N_i (K\in[K_i,K_i+\Delta K])}{\Delta K},
\end{equation}
where $dN/dK_i$ represents the spectrum evaluated for the $i$-th bin with kinetic energy between $[K_i,K_i+\Delta K]$.
Finally in the tabulation of the results we report kinetic energy in terms of $\log_{10}(x)$, where $x = K/M_{\rm{DM}}$.

\end{document}